%% file: 0main.tex
\documentclass[10pt,conference]{IEEEtran}  % for IEEE conference
\IEEEoverridecommandlockouts

\usepackage[ruled,linesnumbered]{algorithm2e} % algo
\usepackage{amsmath,amssymb,amsfonts,bm} % equation
\usepackage{tabularx,booktabs,multirow} % 表格:控制宽度/表格线/多行
\usepackage{graphicx,subcaption,wrapfig} % figure, sub-figures, wrap figures
\usepackage{cite} % 参考文献列表排版
\usepackage[numbers,sort&compress]{natbib}
\usepackage{textcomp} % 额外的文本符号
\usepackage{xcolor}
\usepackage{placeins} %禁止浮动
\newtheorem{problem}{Problem}
\def\BibTeX{{\rm B\kern-.05em{\sc i\kern-.025em b}\kern-.08em
    T\kern-.1667em\lower.7ex\hbox{E}\kern-.125emX}}  % for IEEE

\begin{document}

\title{Enhancing the Trainability of Variational Quantum Circuits with Regularization Strategies}
%\title{Conference Paper Title*\\
%}  % for IEEE
%\author{Anonymous}

\author{\IEEEauthorblockN{
Jun Zhuang*\textsuperscript{1}  \ \
Jack Cunningham\textsuperscript{1}  \ \
Chaowen Guan*\textsuperscript{2}
}
\IEEEauthorblockA{
\textsuperscript{1}Boise State University, ID, USA.
\textsuperscript{2}University of Cincinnati, OH, USA.\\
{*Corresponding authors: junzhuang@boisestate.edu, guance@ucmail.uc.edu}
%\vspace{-0.3in}
}
}

\maketitle  % for IEEE

\begin{abstract}
In the era of noisy intermediate-scale quantum (NISQ), variational quantum circuits (VQCs) have been widely applied in various domains, demonstrating the potential advantages of quantum circuits over classical models. Similar to classic models, VQCs can be optimized by various gradient-based methods. However, the optimization may get stuck in barren plateaus initially or trapped in saddle points during training. These gradient-related issues can severely impact the trainability of VQCs. In this work, we propose a strategy that regularizes model parameters with prior knowledge of the training data and Gaussian noise diffusion. We conduct ablation studies to verify the effectiveness of our strategy across four public datasets and demonstrate that our method can improve the trainability of VQCs against the above-mentioned gradient issues.
\end{abstract}

%\keywords{Barren Plateau; VQA; Survey}  % for ACM
\begin{IEEEkeywords}
Variational Quantum Circuits, Barren Plateau, Regularization, Gaussian Noise Diffusion
\end{IEEEkeywords}  % for IEEE

%\maketitle  % for ACM

%\section{Introduction}
\input{1intro}

%\section{Related Works}
\input{2rewk}

%\section{Methodology}
\input{3method}

%\section{Experiments}
\input{4exp}

\section{Conclusion}
\label{sec:conclusion}
In this study, we propose a regularization strategy integrated with two mechanisms to improve the trainability of variational quantum circuits (VQCs). First, we leverage prior knowledge of the train data to initialize VQCs' model parameters for mitigating barren plateau issues. Second, we regularize the model parameters by diffusing Gaussian noise along each training epoch to avoid the training being trapped in saddle points. In the experiment, we conduct ablation studies to verify the effectiveness of our proposed methods across four public datasets. Experimental results demonstrate that, after incorporating our proposed mechanisms, the gradient variance remains at a higher level as the model scales up, compared to classical or state-of-the-art mitigation strategies.

\section{Limitations and Future Directions}
\label{sec:limitation}
In this study, we empirically verify the effectiveness of our proposed method. However, our method may fail to perform robustly due to the following limitations.
First, we assume that the dataset follows a well-known distribution, so we could regularize the initial distribution with prior knowledge of the training data. However, in real-life scenarios, data distributions may be more complex. This complexity may result in the failure to capture the true data distribution during initialization.
Second, we assume that the distribution of the training data remains static during training. Based on this assumption, our method may be unable to adapt to the distribution shift since we predetermine the initial distribution of model parameters and the diffusion rate.

In the future, for the first limitation, we can employ non-parametric Bayesian approaches to capture the complex data distribution. To address the second limitation, we could handle the distribution-shift problem via detection-based methods (for detecting the shift) or adaptation-based methods (for adaptively updating the hyperparameters).

%\begin{acks}
%\end{acks}

%\appendix{\huge \textbf{Appendix}}
\appendix
\input{5appendix}

%%
%% The next two lines define the bibliography style to be used, and the bibliography file.
%\bibliographystyle{ACM-Reference-Format}  % for ACM
\bibliographystyle{IEEEtran}  % for IEEE
\bibliography{2reference}
\clearpage

\end{document}

%% file: 1intro.tex
\section{Introduction}
\label{sec:intro}

In recent years, there have been significant advancements in quantum information, particularly with the advent of noisy intermediate-scale quantum (NISQ) devices~\cite{preskill2018quantum}. Within this research landscape, variational quantum circuits (VQCs) have been widely applied in various domains, such as quantum machine learning~\cite{zhang2024generative, zhuang2025large}, quantum physics~\cite{chen2017multi, chen2022stable}, and quantum hardware architecture~\cite{zhan2023quantum, zhan2023optimizing}. Typical VQCs are trainable random parameterized quantum circuits or classical-quantum hybrid models~\cite{li2021vsql}. Similar to classic models, VQCs can be optimized by various gradient-based approaches, such as Adam~\cite{kingma2014adam}.
However, optimization processes may encounter some gradient issues. Primarily, the initialization of VQCs might get stuck in a barren plateau landscape. McClean et al.~\cite{McClean2018landscapes} first systematically studied the barren plateau (BP) issues and verified that the gradient variance will exponentially decrease as the model size increases when the VQCs satisfy the assumption of the 2-design Haar distribution. Under these circumstances, most gradient-based approaches would fail. Additionally, the optimization may be trapped in saddle points during training~\cite{ge2015escaping, jin2017escape}. Both gradient issues can significantly weaken the trainability of VQCs.

Extensive research has focused on addressing the barren plateau problem, with initialization-based strategies proving effective by initializing VQC parameters with diverse distributions~\cite{sack2022avoiding}. For instance, Zhang et al.~\cite{zhang2022escaping} verify the effectiveness of Gaussian initialization on VQCs with a well-designed variance. Nevertheless, most initialization strategies neglect the impact of real data distribution. To address this oversight, Prince argues that applying the data posterior to model parameters could provide more robust performance~\cite{prince2023understanding}. However, posterior estimation on complex models may not be practical due to high computational overhead. To overcome the above drawbacks, we leverage prior knowledge of the training data to regularize the initial distribution of model parameters. Our intuition for this regularization is that incorporating prior knowledge in initialization can better shape the initial distribution of model parameters to some extent, thus helping alleviate barren plateaus.

\begin{figure}[t]
  \begin{subfigure}{1\linewidth}
    \centering
    \includegraphics[width=0.49\textwidth]{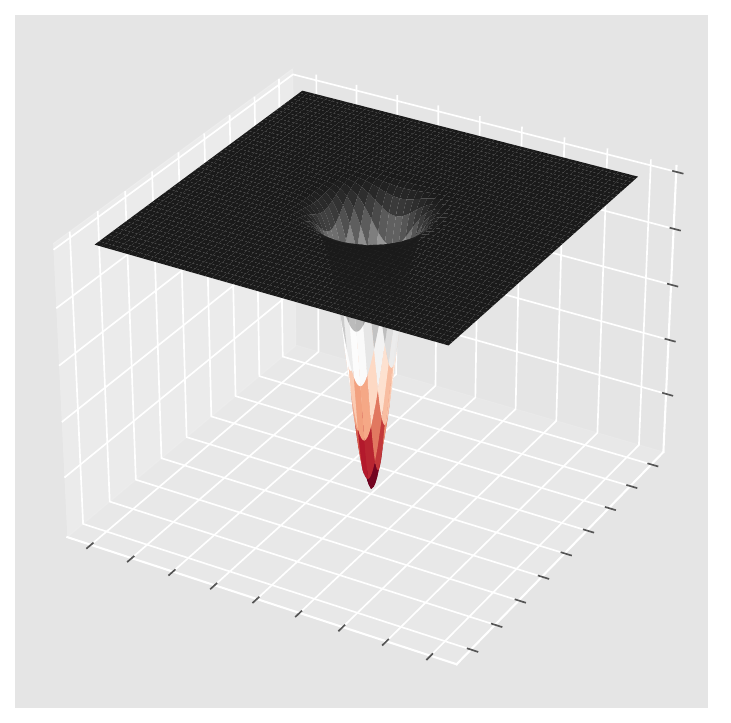}
    \includegraphics[width=0.49\textwidth]{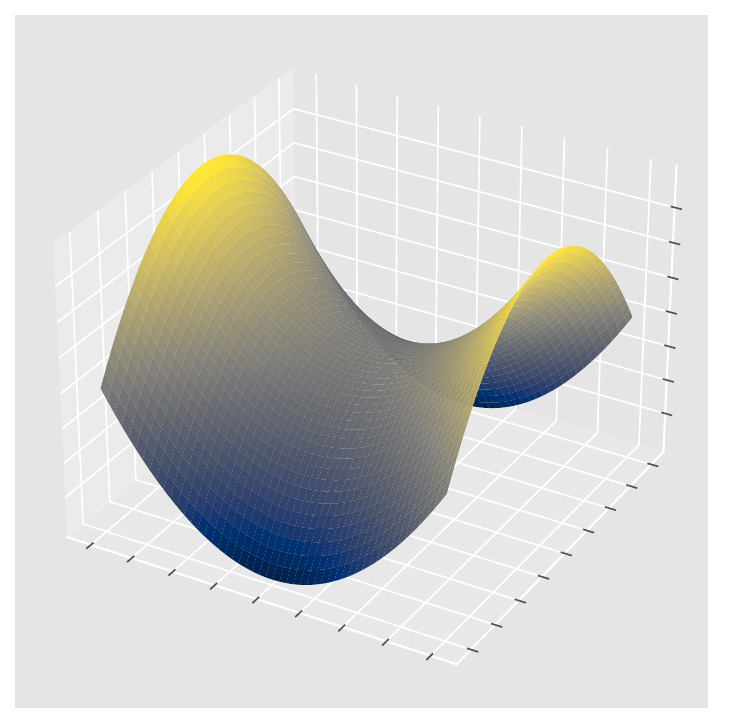}
  \end{subfigure}
\caption{Examples of loss landscapes on the barren plateau (left side) and the saddle point (right side).}
\label{fig:ex_bp_sp}
\end{figure}

Besides mitigating barren plateau issues via initialization-based strategies, some studies improve the trainability by adding noise to model parameters to avoid being trapped in saddle points during training ~\cite{kulshrestha2022beinit}. However, adding excessive noise may potentially undermine model performance~\cite{zhuang2023robust}. Inspired by DDPM~\cite{ho2020denoising}, which models the data distribution by iteratively diffusing Gaussian noise during data generation, we gradually diffuse Gaussian noise on model parameters during training. The intuition behind noise diffusion is that as training converges, the model will gradually perform better, thus requiring slightly noise perturbations on model parameters.
To better understand the above-mentioned two issues, we present examples of their loss landscapes in Figure~\ref{fig:ex_bp_sp}.

By integrating the above two mechanisms, in this study, we propose a regularization strategy to improve the trainability of VQCs. In our proposed method, we first leverage prior knowledge of the training data to regularize the initial distribution of model parameters, and further diffuse Gaussian noise on the parameters along each training iteration.
In experiments, we conduct ablation studies to examine the effectiveness of our proposed methods.
First, we validate that leveraging prior knowledge of the train data can effectively regularize three prevalent initial distributions of model parameters and yield superior mitigation on barren plateau issues. 
Furthermore, we affirm that diffusing Gaussian noise to model parameters during training can effectively increase volatility to avoid being trapped in saddle points while adequately alleviating the degradation of gradient variance on three methods. Last, we analyze the key hyperparameter, max diffusion rate ($dr_{max}$), on Normal distribution as an example. Extensive results demonstrate the effectiveness of our proposed regularization strategy over four public datasets.
Overall, our contributions to this study can be summarized as follows:
\begin{itemize}
  \item We propose a strategy that regularizes model parameters with prior knowledge of the train data and diffused Gaussian noise for improving the trainability of VQCs.
  \item We conduct extensive experiments to verify the effectiveness of our proposed method across four public datasets.
\end{itemize}

%% file: 2rewk.tex
\section{Related Work}
\label{sec:rewk}
In this section, we introduce the related work about barren plateaus and saddle points in the following paragraphs.

\paragraph{Barren Plateau}
McClean et al.~\cite{McClean2018landscapes} first investigated barren plateau (BP) phenomena and demonstrated that under the assumption of the 2-design Haar distribution, gradient variance in VQCs will exponentially decrease to zero during training as the model size increases. In recent years, extensive studies have been devoted to mitigating barren plateau issues in VQCs. Recent studies~\cite{cunningham2025investigating} categorize these efforts as 
initialization-based strategies~\cite{Friedrich2022avoiding, mele2022avoiding, grimsley2023adaptive, liu2023mitigating, park2024hamiltonian},
optimization-based strategies~\cite{haug2021optimal, wu2021mitigating, liu2024mitigating, sannia2023engineered, gharibyan2023hierarchical, sciorilli2024towards, falla2024graph},
model-based strategies~\cite{selvarajan2023dimensionality, zhang2022quark, shin2024layerwise},
and measurement-based strategies~\cite{rappaport2023measurement}.
First, {\bf initialization-based strategies} mainly aim to initialize model parameters with different distributions. Within this category, Grant et al.~\cite{grant2019initialization} propose an identity block strategy that can initialize the VQCs as a sequence of blocks of identity operators. Sauvage et al.~\cite{sauvage2021flip} propose a flexible initializer (FLIP) for arbitrarily sized VQCs. Kulshrestha et al.~\cite{kulshrestha2022beinit} initialize the circuit parameters from a beta distribution. Zhang et al.~\cite{zhang2022escaping} verify that applying Gaussian initialization with well-designed variance can mitigate barren plateau issues.
Second, {\bf optimization-based strategies} primarily improve the efficiency of optimization while mitigating barren plateaus as well. For example, Ostaszewski et al.~\cite{ostaszewski2021structure} propose a new method for effectively optimizing VQCs' structure and parameters with lower computational overhead. Suzuki et al.~\cite{suzuki2021normalized} propose a new normalized gradient descent (NGD) method that can converge faster than the naive NGD-based method. Heyraud et al.~\cite{Heyraud2023estimation} design an efficient method to compute the gradient for a wide range of VQCs.
Third, {\bf model-based strategies} address barren plateau problems via various model architectures. For example, Li et al.~\cite{li2021vsql} propose a hybrid quantum-classical framework, namely VSQL, to avoid barren plateaus. Concurrently, Bharti and Haug~\cite{Bharti2021simulator} propose a hybrid quantum-classical algorithm for dynamically simulating quantum circuits. Du et al.~\cite{du2022quantum} design an efficient search scheme, QAS, to automatically seek a near-optimum during VQCs' training. Tüysüz et al.~\cite{tuysuz2023classical} propose a model to divide the VQCs into multiple sub-circuits to avoid barren plateaus. Kashif et al.~\cite{Kashif2024resQNets} introduce residual quantum neural networks (ResQNets) by splitting QNN architectures into multiple quantum nodes.
Last but not least, {\bf measurement-based strategies} investigate the barren plateau landscape in hybrid variational quantum circuits from the perspective of measurement~\cite{rappaport2023measurement}.

\paragraph{Saddle Point}
In recent years, saddle point issues have posed a considerable challenge to numerical optimization~\cite{benzi2005numerical}. Conventional optimization methods primarily focus on local extrema, but modern research has revealed that high-dimensional non-convex loss functions often contain numerous saddle points rather than local minima~\cite{dauphin2014identifying}. In classical machine learning, gradient-based optimization approaches have been observed to effectively circumvent high-order saddle points to some extent, while the efficiency of escaping low-order saddle points depends on noise or specific optimization strategies~\cite{jin2017escape}.

Extended from classical to quantum machine learning, saddle point issues are also critical, particularly in the quantum models, such as variational quantum circuits (VQCs)~\cite{zhang2021quantum}. 
Sun et al.~\cite{sun2015measuring} propose metrics to quantify the distance from saddle points in quantum control landscapes.
Recent studies reveal that inherent stochastic noise in VQCs naturally helps training avoid being trapped in saddle points, providing convergence guarantees and validating the concept through simulations and quantum hardware experiments~\cite{liu2022stochastic}. To further leverage the noise, Kulshrestha et al.~\cite{kulshrestha2022beinit} manually inject noise during training VQCs. Overall, research on saddle points continues to evolve, and developing efficient strategies to mitigate their impact—especially in high-dimensional quantum optimization problems—remains an open challenge.

%% file: 3method.tex
\section{Methodology}
\label{sec:method}
In this section, we first introduce the background about variational quantum circuits, barren plateaus, saddle points, and clearly state the problem that we aim to solve. Besides, we describe our proposed strategy with two mechanisms and further explain our training procedure with an analysis of time and space complexity.

\begin{figure*}[t]
\centering
  \includegraphics[width=\linewidth]{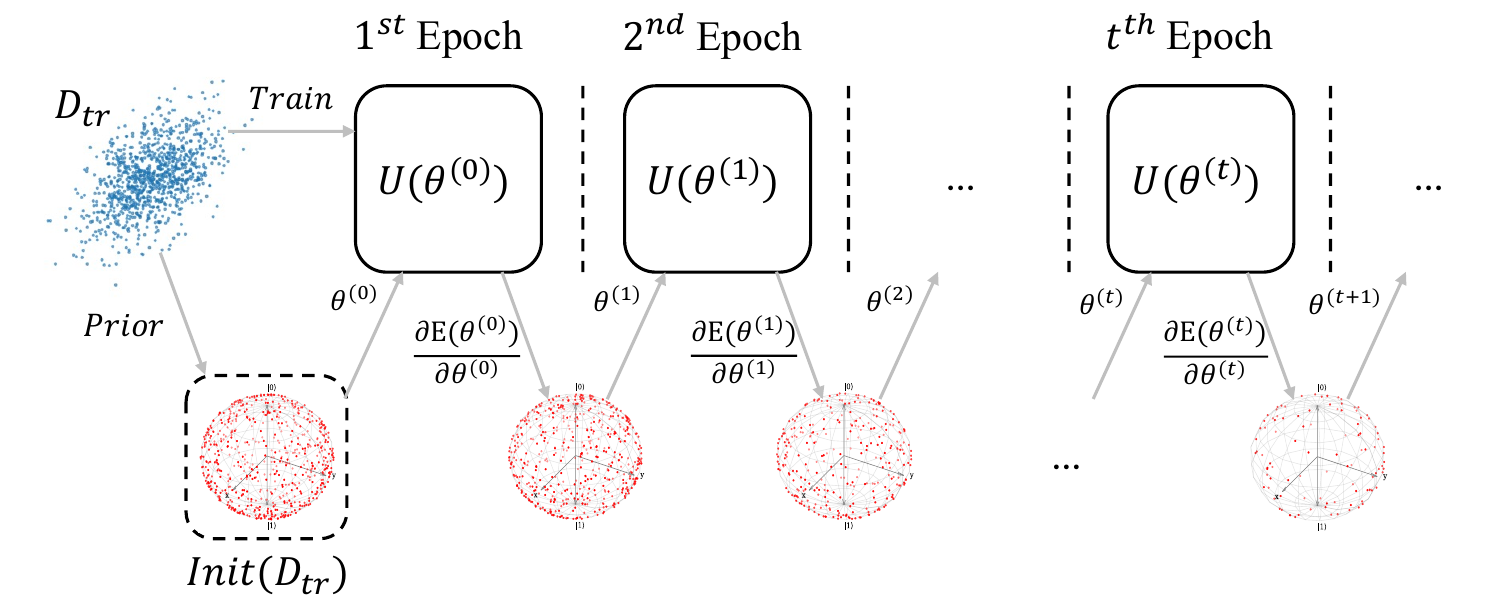}
  \caption{The overall process of our proposed strategy. Given a training data $D_{tr}$, we first initialize the model parameters with the prior knowledge of $D_{tr}$ as $\theta^{(0)}$ and feed both $D_{tr}$ and $\theta^{(0)}$ to the VQC $U(\cdot)$ for iterative training. In each iteration, let's say in the $t^{th}$ iteration, we update $\theta^{(t)}$ with the gradient $\frac{\partial E(\theta^{(t)})}{\partial \theta^{(t)}}$ via a gradient-based approach and further diffuse Gaussian noise on model parameters $\theta^{(t+1)}$ for the next iteration.}
\label{fig:train}
\end{figure*}

\subsection{Preliminary}
\paragraph{Variational Quantum Circuits}
Typical VQCs consist of a finite sequence of unitary gates $U(\theta)$ parameterized by $\theta \in \mathbb{R}^{NRL}$, where $N$, $R$, and $L$ denote the number of qubits, rotation gates, and layers. $U(\theta)$ can be formulated as:
%\vspace{-1mm}
\begin{equation}
    U(\theta) = U(\theta_1, ..., \theta_L) = \prod_{l=1}^{L} U_l(\theta_l)W_l,
\label{eqn:vqc}
\end{equation}
%\vspace{-1mm}
where $U_l(\theta_l) = e^{-i\theta_lV_l}$, $V_l$ is a Hermitian operator, and $W_l$ is unitary operator that doesn't depend on $\theta_l \in \mathbb{R}^{NR}$.

VQCs can be optimized by gradient-based methods. To achieve this goal, we first define the loss function $E(\theta)$ of $U(\theta)$ as the expectation over the Hermitian operator $H$:
\begin{equation}
    E(\theta) = \langle0| U(\theta)^{\dagger} H U(\theta) |0\rangle.
\label{eqn:loss_fn}
\end{equation}

Given the loss function $E(\theta)$, we can further compute its gradient by the following formula:
\begin{equation}
    \partial_k E \equiv \frac{\partial E(\theta)}{\partial \theta_k} = i\langle0| U_{-}^{\dagger} \left[ V_k, U_{+}^{\dagger} H U_{+} \right] U_{-} |0\rangle,
\label{eqn:gradient}
\end{equation}
where $U_{-} \equiv \prod_{l=0}^{k-1} U_l(\theta_l)W_l$, $U_{+} \equiv \prod_{l=k}^{L} U_l(\theta_l)W_l$. Also, $U(\theta)$ is sufficiently random s.t. both $U_{-}$ and $U_{+}$ (or either one) are independent and match the Haar distribution up to the second moment.

\begin{figure}[h]
\centering
  \includegraphics[width=\linewidth]{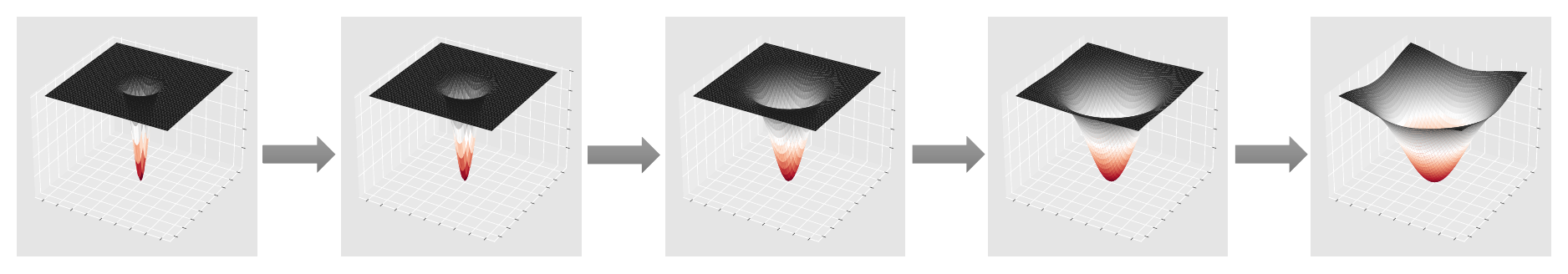}
  \caption{Example of a barren plateau mitigation process from left to right. We present a loss landscape of BPs in the left-most sub-figure while displaying a recovered loss landscape in the right-most sub-figure.}
\label{fig:bp}
\end{figure}

\paragraph{Barren Plateaus}
McClean et al.~\cite{McClean2018landscapes} first investigated Barren Plateau issues in VQCs. They conduct experiments on random VQCs to verify that gradient variance $Var[\partial_k E]$ will exponentially decrease as the number of qubits $N$ increases when the VQCs, such as $U_{-}$ or $U_{+}$, match 2-design Haar distribution. This relationship can be approximated as follows:
\begin{equation}
    Var[\partial_k E] \propto 2^{-2N}.
\label{eqn:var}
\end{equation}
The Equation~\ref{eqn:var} indicates that in most cases, $Var[\partial_k E]$ will approximate zero when the number of qubits $N$ is very large. In other words, most gradient-based approaches will fail to train VQCs under the above circumstances, which is visualized in the left-most sub-figure in Figure~\ref{fig:bp}. In this study, we aim to mitigate the BPs, whose recovery process is presented in Figure~\ref{fig:bp} as an example.

\paragraph{Saddle Points}
In high-dimensional optimization, saddle points represent a fundamental challenge to efficient convergence in training classical deep learning models. A saddle point is a stationary point where the gradient vanishes, but unlike local minima, it possesses at least one direction in which the function exhibits negative curvature~\cite{pascanu2014saddle}. This gradient issue will lead to suboptimal training due to being trapped in saddle points. Thus, it is crucial to escape saddle points. One effective approach for escaping saddle points involves injecting Gaussian noise into the model parameters during training~\cite{staib2019escaping}. Formally, for model parameter $\theta^{(t)}$ in the $t$-th iteration, the update rule with Gaussian noise can be expressed as follows:
\begin{equation}
  \theta^{(t+1)} \gets \theta^{(t)} - \eta \frac{\partial E(\theta^{(t)})}{\partial \theta^{(t)}} + \epsilon,
\label{eqn:add_noise}
\end{equation}
where $\eta$ denotes the learning rate and $\epsilon \sim \mathcal{N}(0, I)$ denotes the standard Gaussian noise.
%\begin{wrapfigure}{r}{0.4\linewidth}
  % \centering
  % \includegraphics[width=0.8\linewidth]{figures/fig_sp.pdf}
  % \caption{Example of a saddle point (SP).}
  % \label{fig:sp}
%\end{wrapfigure}

\paragraph{Problem Statement}
To clarify our goal, we summarize the problem that we aim to solve in this study as follows.
\begin{problem}
To improve the trainability of VQCs, we aim to simultaneously address both barren plateau and saddle point issues such that the gradient variance of VQCs can be maximized.
\label{prob:statement}
\end{problem}

\subsection{Our Proposed Regularization Strategy}
To improve the trainability of VQCs, in this study, we propose a strategy that regularizes model parameters with i) prior knowledge of the training data in the initialization and ii) Gaussian noise diffusion during optimization. To better illustrate our proposed strategy, we briefly introduce the overall process in Figure~\ref{fig:train}.
In the following subsections, we will introduce these two mechanisms in detail.

\paragraph{Regularization with Prior Knowledge}
Regularizing the model parameters via Bayesian inference is a popular regularization technique~\cite{zhu2014bayesian}. Specifically, the Bayesian approach can regularize the parameters by initializing the model weights using the approximated posterior distributions. This approach regards the model parameters $\theta$ as unknown variables and computes a posterior distribution ${\it P} \left( \theta \mid D \right)$ over $\theta$ given the data $D$ as a condition. According to Bayes' rule, the posterior can be approximated as follows:
\begin{equation}
{\it P} \left( \theta \mid D \right) \propto {\it P} \left( D \mid \theta \right) {\it P} \left( \theta \right),\\
\label{eqn:bayes}
\end{equation}
where ${\it P} \left( D \mid \theta \right)$ denotes the maximum likelihood and ${\it P} \left( \theta \right)$ denotes the prior distribution of model parameters.

\begin{figure}[h]
  \centering
  \includegraphics[width=\linewidth]{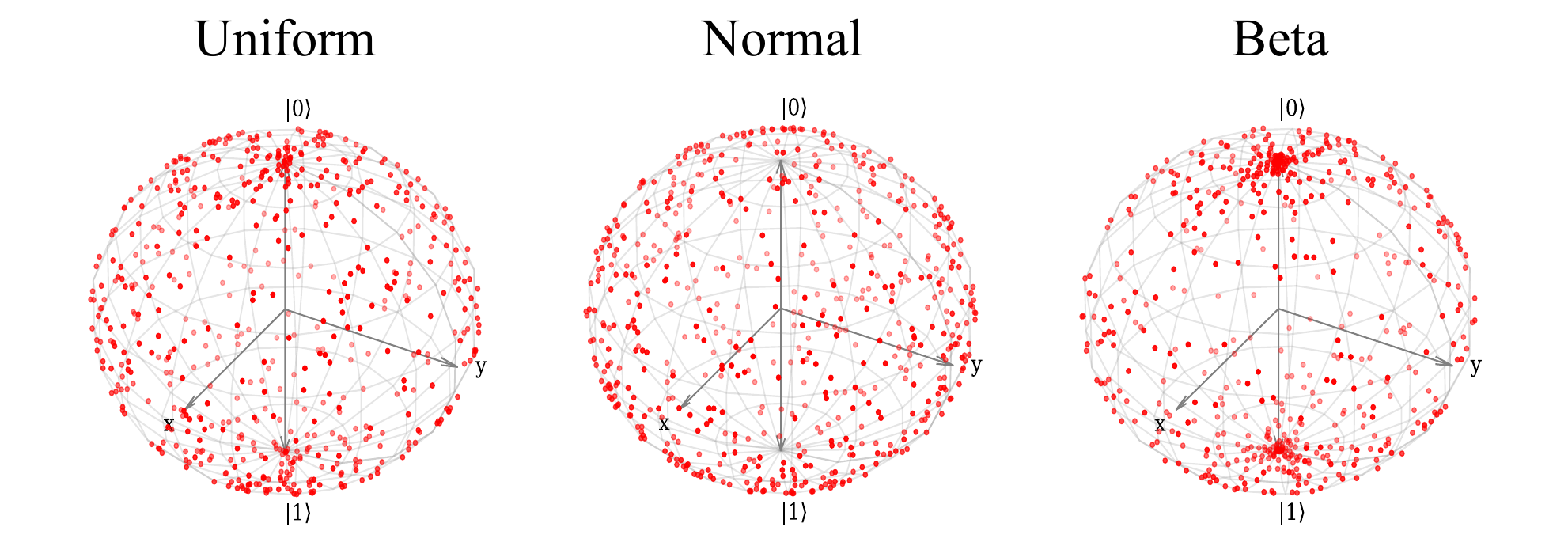}
  \caption{Example of three original distributions for initialization. The red points in this figure represent the initial values of model parameters.}
\label{fig:prior}
\end{figure}

\begin{table}[h]
\centering
\caption{The distinctions of initialization between the original distributions and our distributions on three classic distributions. $D_{min}$ and $D_{max}$ denote the minimum and maximum values of the given data $D$, such as the train data $D_{tr}$ in the study. $\mu_D$, $\sigma_D$, $\alpha_D$, and $\beta_D$ denote the corresponding hyperparameters derived from the data $D$.}
\label{tab:dist}
\begin{tabular}{ccc}
  \toprule
    Distribution & Original & Ours \\
    \midrule
    Uniform & $ \text{Uniform}\left(0, 1\right)$ & $\text{Uniform}\left(D_{min}, D_{max}\right)$ \\
    Normal & $\text{Normal}\left(0, 1\right)$ & $\text{Normal}\left(\mu_D, \sigma_D\right)$ \\
    Beta & $\text{Beta}\left(0.5, 0.5\right)$ & $\text{Beta}\left(\alpha_D, \beta_D\right)$ \\
  \bottomrule
\end{tabular}
\end{table}
Employing posterior as the initial distribution of model parameters can provide a more robust initialization than only using maximum likelihood~\cite{prince2023understanding}. However, this approach has a significant drawback. For complex models, such as deep neural networks, it is not practical to compute the full probability distribution over model parameters due to high computational costs. To overcome this drawback, we simplify the process by considering the prior distribution of the given data $D$, such as the train data $D_{tr}$, as the initial distribution of model parameters. Our intuition is that utilizing such prior knowledge in initialization is equivalent to setting a constraint on the search space, thereby providing a robust initial optimization landscape against barren plateau issues.
In Table~\ref{tab:dist}, we present the distinctions of three classic initial distributions between using original distributions (``Original'') and considering prior knowledge (``Ours'') as examples.
We also visualize the initial distributions as an example to better demonstrate their distinctions among the three original distributions in Figure~\ref{fig:prior}.

\paragraph{Regularization with Gaussian Noise Diffusion}
Besides the Bayesian approach, adding noise is another popular approach for regularization~\cite{noh2017regularizing}. For example, BeInit~\cite{kulshrestha2022beinit} mitigates the degradation of gradient variance by adding Gaussian noise in the model parameters during training. However, adding too much noise will inevitably weaken the model's performance~\cite{zhuang2023robust}. 
Inspired by DDPM~\cite{ho2020denoising}, which models better data distributions via an iterative Gaussian diffusion process, we iteratively diffuse the Gaussian noise on model weights during training. In the $t$-th iteration, we gradually diffuse the standard Gaussian noise $\epsilon \sim \mathcal{N}(0, I)$ to the diffused model parameters with a decreasing diffusion rate $\gamma$ as follows:
\begin{equation}
\left\{
  \begin{array}{lr}
    \overline{\theta^{(t)}} = \sqrt{\Gamma^{(t)}}\theta^{(t)}, & \\
    \overline{\epsilon} = \sqrt{( 1 - \Gamma^{(t)})}\epsilon,
  \end{array}
\right.
\label{eqn:diffusion}
\end{equation}
where $\overline{\theta^{(t)}}$ and $\overline{\epsilon}$ denote the diffused parameters in the $t$-th iteration and diffused Gaussian noise; $\Gamma^{(t)} = \prod_{i=0}^{t} \gamma^{(i)}$ is the accumulated production of previous diffusion rates in the $t$-th iteration. The $\gamma$ linearly decreases with each iteration.

In each iteration, we apply the diffusion process to model parameters after back-propagation. The diffused parameters will be used in the next iteration. This diffusion process can be formulated as follows:
\begin{equation}
\theta^{(t+1)} = \overline{\theta^{(t)}} + \overline{\epsilon}.
\label{eqn:difproc}
\end{equation}

\begin{figure}[h]
  \centering
  \includegraphics[width=\linewidth]{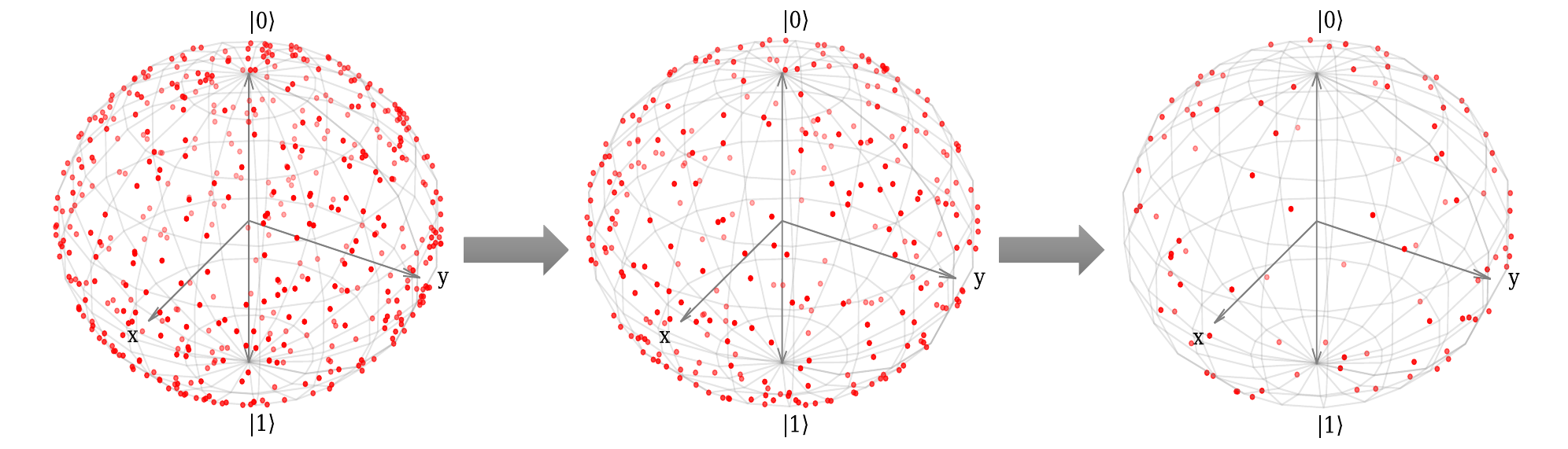}
  \caption{The process of diffusing Gaussian noise. The red points in this figure represent Gaussian noise added as regularization.}
\label{fig:ns_diff}
\end{figure}

As the number of iterations increases, the training will gradually converge, resulting in better model performance. Therefore, the model may require lower noise perturbations for regularization. Based on this intuition, we progressively diffuse the Gaussian noise on the model parameters as the optimization proceeds. We visualize the noise diffusion process in Figure~\ref{fig:ns_diff} as an illustration example.

\begin{algorithm}[h]
\DontPrintSemicolon
\KwIn{Variational quantum circuits $U(\cdot)$, Train data $D_{tr}$, Learning rate, $\eta$, Train epochs $T$}
$\theta^{(0)} \gets Init \left( D_{tr} \right)$;\\
Compute $\Gamma = [\Gamma^{(0)}, \Gamma^{(1)}, ..., \Gamma^{(T-1)}]$;\\
\For{$t = 0 \gets T$}{
 $\theta^{(t)} \gets \theta^{(t)} - \eta \frac{\partial E(\theta^{(t)})}{\partial \theta^{(t)}}$;\\
 $\overline{\theta^{(t)}} \gets \sqrt{\Gamma^{(t)}}\theta^{(t)}$;\\
 $\overline{\epsilon} \gets \sqrt{( 1 - \Gamma^{(t)})}\epsilon$, where $\epsilon \sim \mathcal{N}(0, I)$;\\
 $\theta^{(t+1)} \gets \overline{\theta^{(t)}} + \overline{\epsilon}$;\\
}
\caption{{\sc Our} training procedure.}
\label{algo:train}
\end{algorithm}

\begin{figure*}[t]
  %\hfill
  \begin{subfigure}{0.5\linewidth}
    \centering  % include the 1st and 2nd images
    \includegraphics[width=0.49\textwidth]{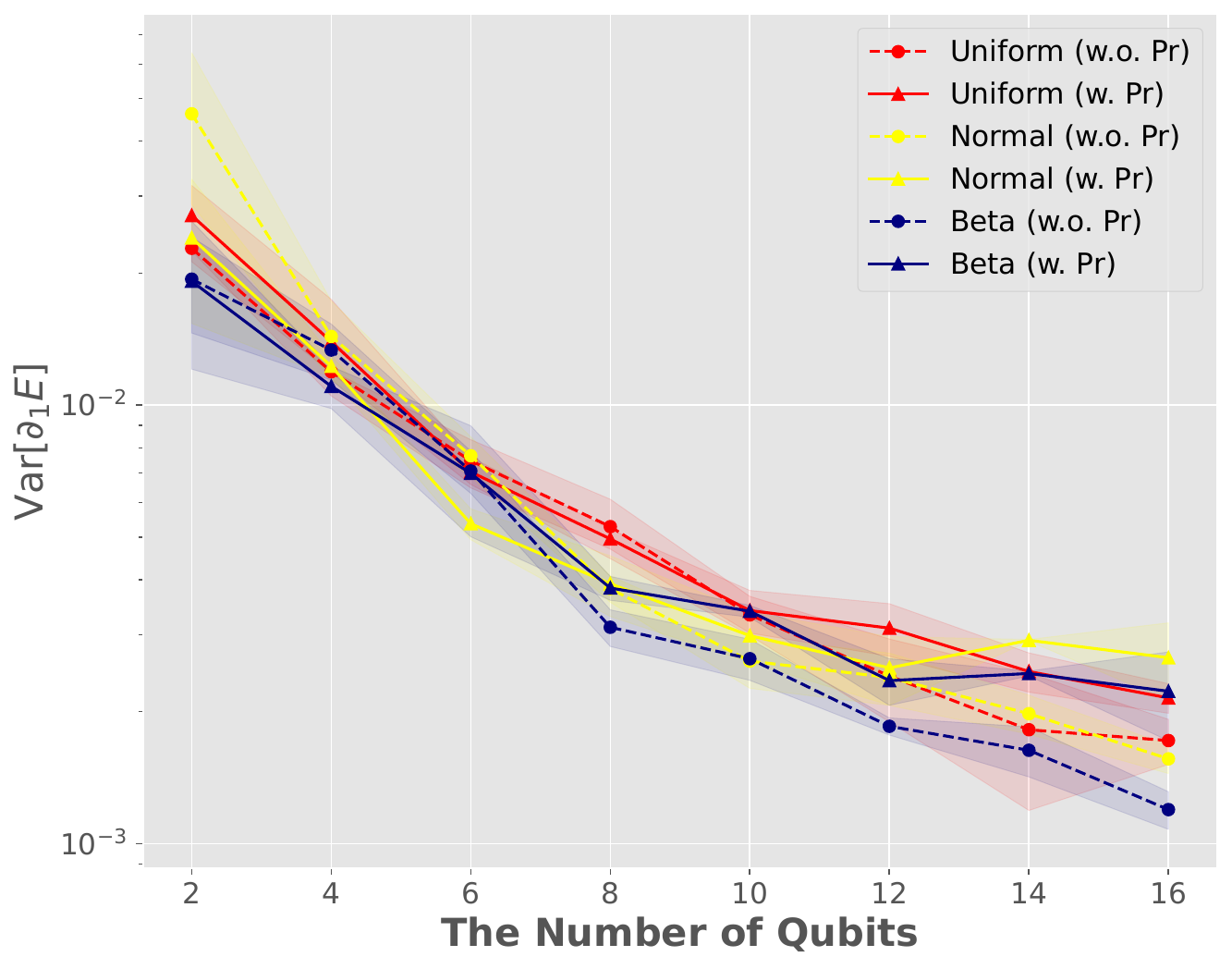}
    \includegraphics[width=0.49\textwidth]{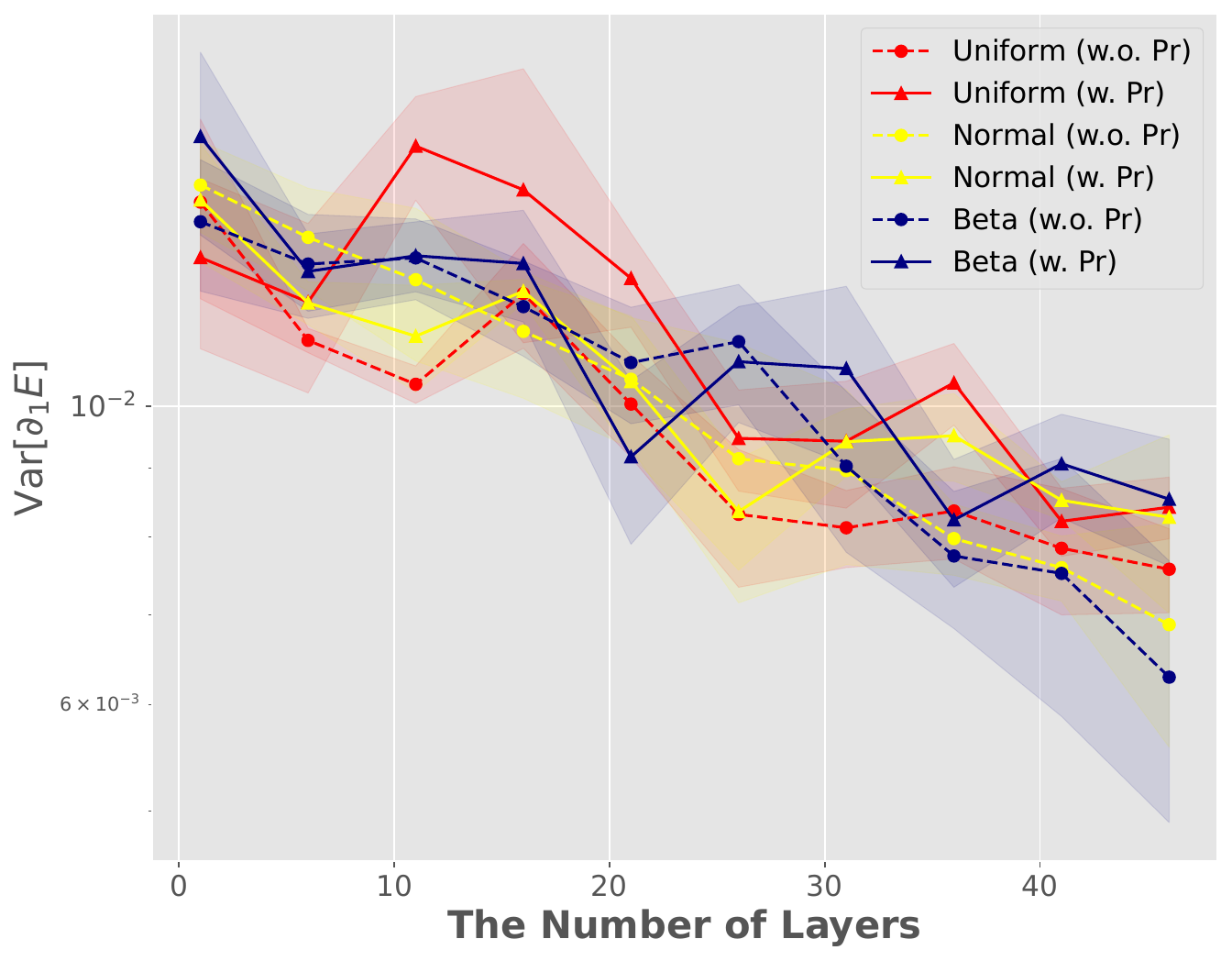}
    \caption{Iris}
  \end{subfigure}
  %\hfill
  \begin{subfigure}{0.5\linewidth}
    \centering  % include the 3rd and 4th images
    \includegraphics[width=0.49\textwidth]{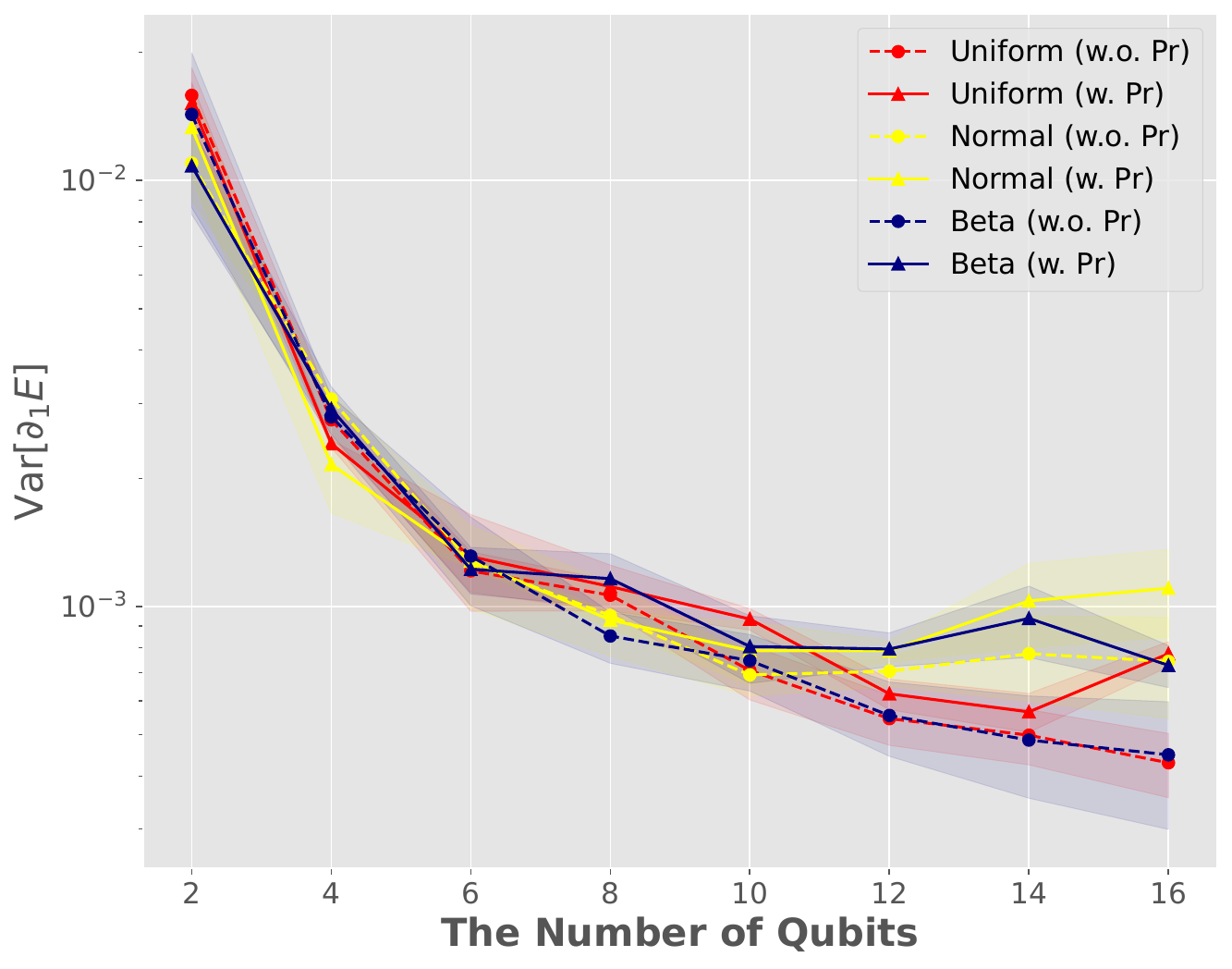}
    \includegraphics[width=0.49\textwidth]{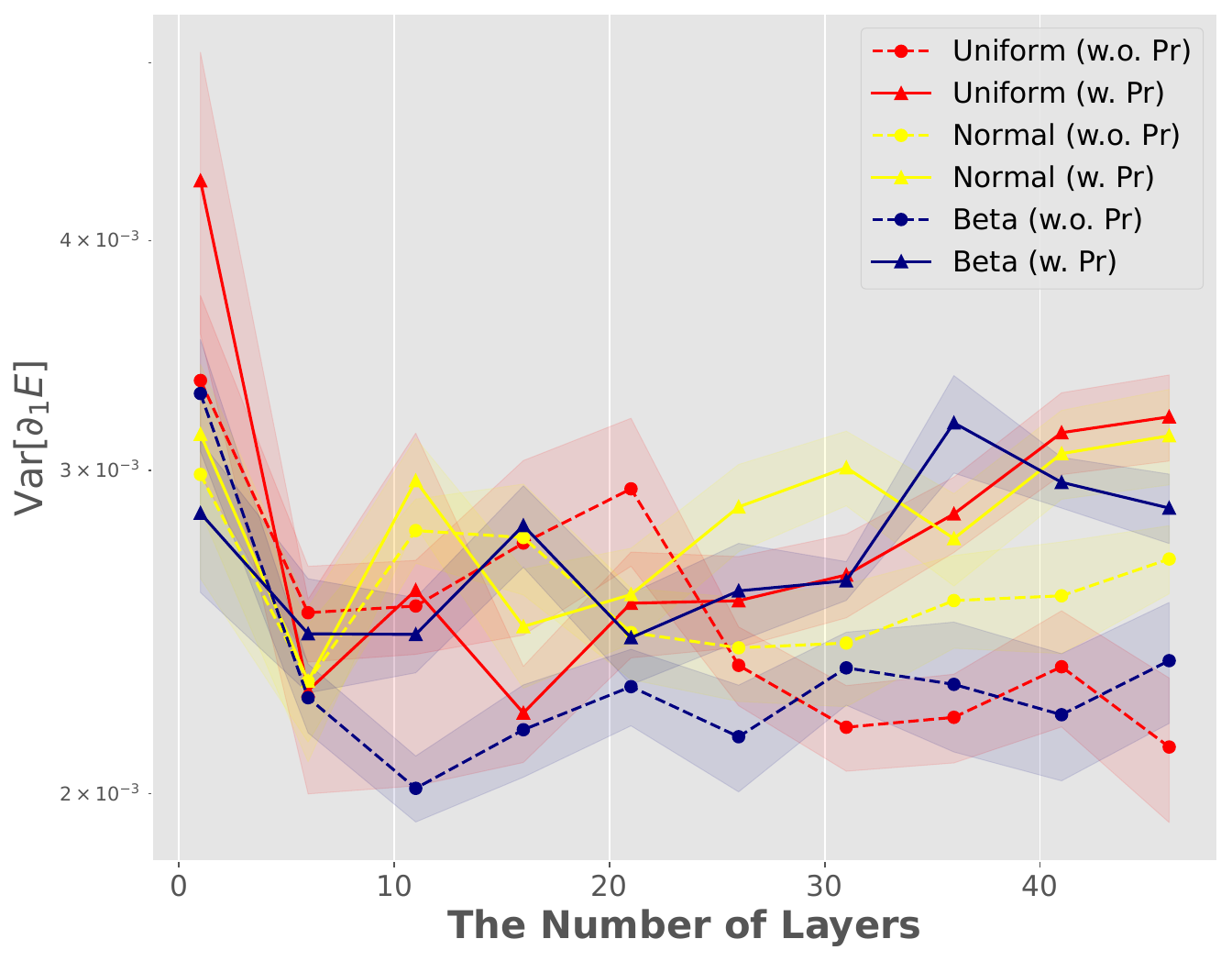}
    \caption{Wine}
  \end{subfigure}
  %\hfill
  \begin{subfigure}{0.5\linewidth}
    \centering  % include the 5th and 6th images
    \includegraphics[width=0.49\textwidth]{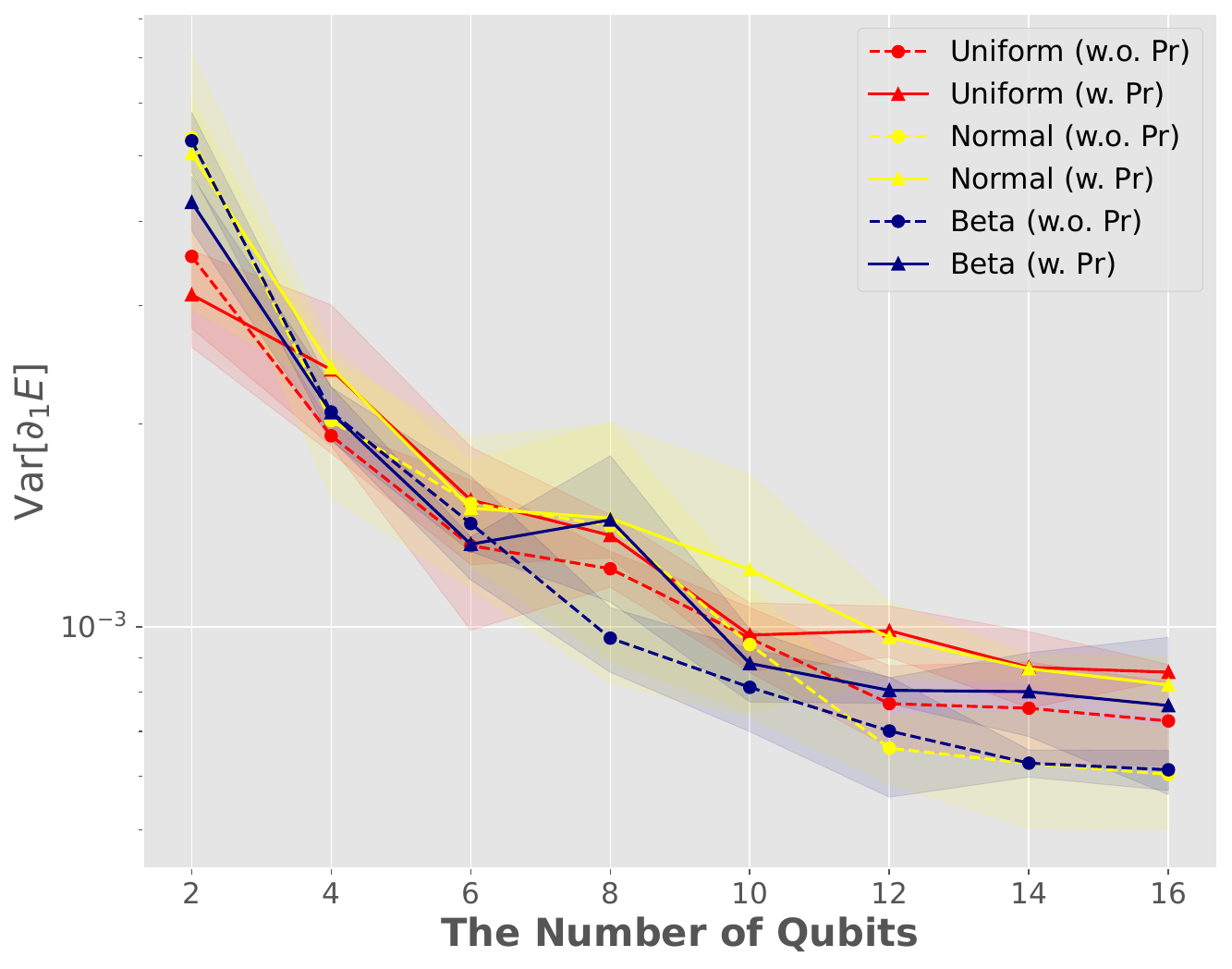}
    \includegraphics[width=0.49\textwidth]{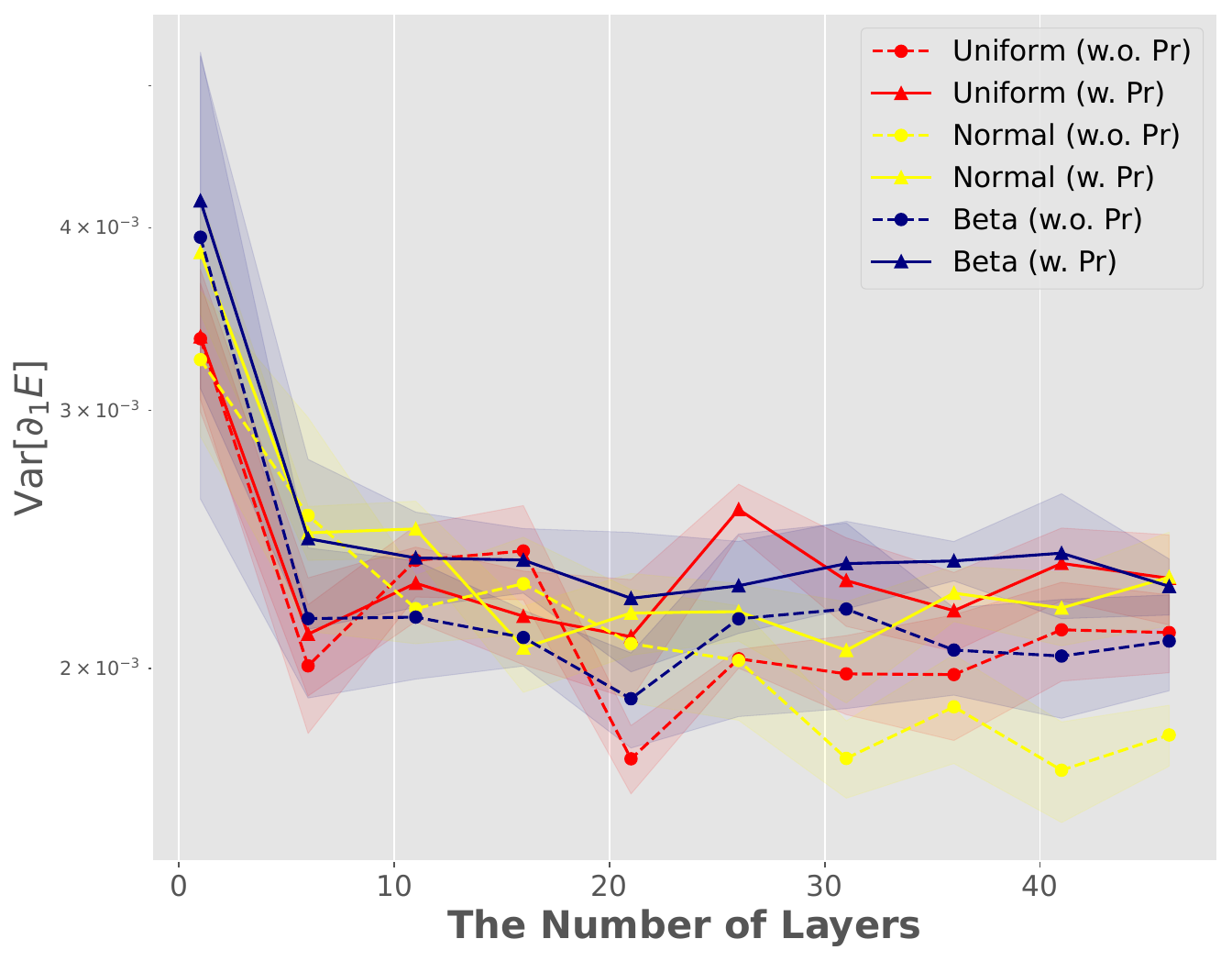}
    \caption{Titanic}
  \end{subfigure}
  %\hfill
  \begin{subfigure}{0.5\linewidth}
    \centering  % include the 7th and 8th images
    \includegraphics[width=0.49\textwidth]{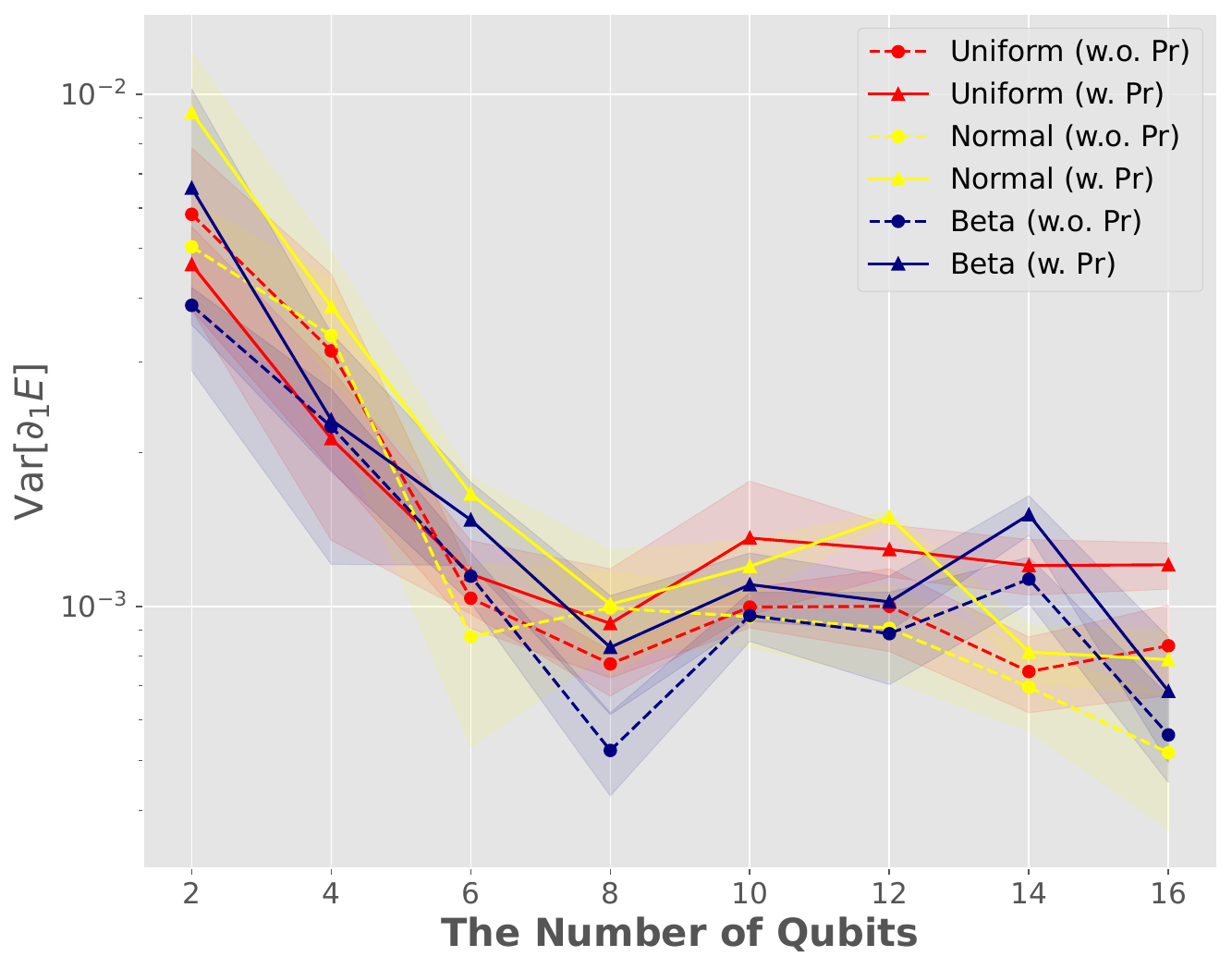}
    \includegraphics[width=0.49\textwidth]{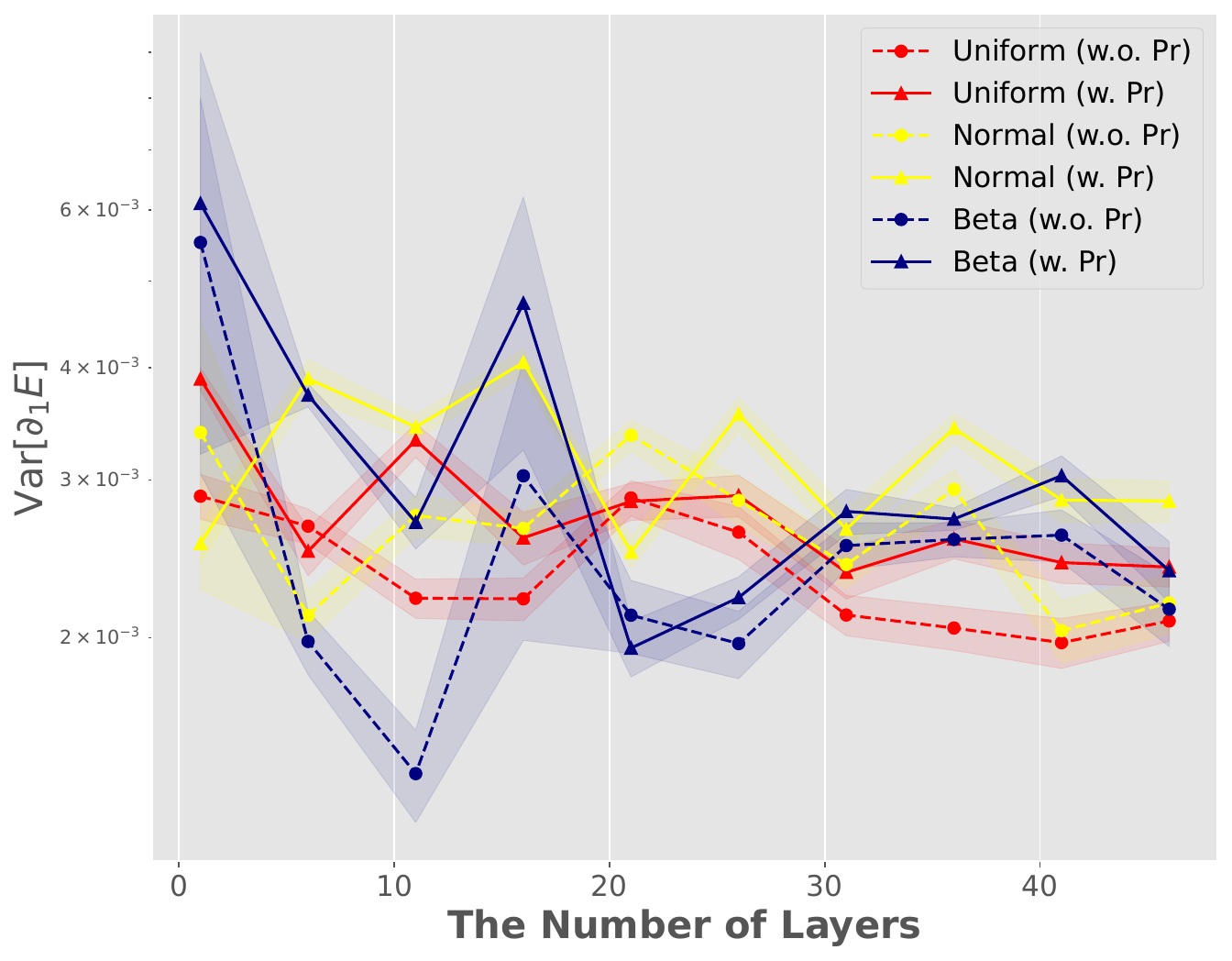}
    \caption{MNIST}
  \end{subfigure}
\caption{Investigation of the mechanism that leverages prior knowledge of the train data on three classic initialization methods. ``w. Pr'' and ``w.o. Pr'' denote whether or not we apply prior knowledge.}
\label{fig:pr_exp}
\end{figure*}

\paragraph{The Training Procedure}
As presented in Algo.~\ref{algo:train}, we first initialize the model parameters $\theta^{(0)}$ with prior knowledge of the train data $D_{tr}$ one time ({\bf line 1}), and then compute the hyperparameter $\Gamma$, for each train step ({\bf line 2}).
After initialization, we train the variational quantum circuit $U(\cdot)$ with $T$ epochs ({\bf line 3-8}). In the $t$-th iteration of the train loop, we update the model parameters $\theta^{(t)}$ via optimization approaches, such as gradient descent, with a learning rate $\eta$, where the gradient denotes $\frac{\partial E(\theta^{(t)})}{\partial \theta^{(t)}}$ ({\bf line 4}). After updating the model parameters $\theta^{(t)}$, we apply diffusion to $\theta^{(t)}$ and Gaussian noise $\epsilon$ with $\Gamma^{(t)}$ using Equation~\ref{eqn:diffusion} ({\bf line 5-6}) and further update $\theta^{(t+1)}$ using Equation~\ref{eqn:difproc} for the next iteration ({\bf line 7}).

\paragraph{Analysis of Time and Space Complexity}
We propose two mechanisms for regularization in the training procedure. Regularizing the initial distribution with prior knowledge of the train data only implements once in {\bf line 1} and thus takes $\mathcal{O}(1)$. On the other hand, diffusing Gaussian noise to the model parameters $\theta^{(t)}$ takes constant time $\mathcal{O}(3)$ in each iteration ({\bf line 5-7}). So, the total {\bf time complexity} for $T$ train loops would be $\mathcal{O}(T+3) \approx \mathcal{O}(T)$.
For the space complexity, initialization with prior knowledge does not take extra space, whereas diffusing Gaussian noise does require extra intermediate spaces for $\overline{\theta^{(t)}}$ and $\overline{\epsilon}$ in each iteration, but these spaces are constant and will be released after each iteration. Thus, the total {\bf space complexity} is still $\mathcal{O}(\theta)$.
Overall, our proposed regularization methods will not theoretically increase time and space complexity.

%% file: 4exp.tex
\section{Experiments}
\label{sec:exp}
In this section, we first introduce the experimental settings. Second, we present ablation studies to validate the effectiveness of two proposed mechanisms. At last, we present the optimal hyperparameters for our method.

\begin{figure*}[h]
  %\hfill
  \begin{subfigure}{0.5\linewidth}
    \centering  % include the 1st and 2nd images
    \includegraphics[width=0.49\textwidth]{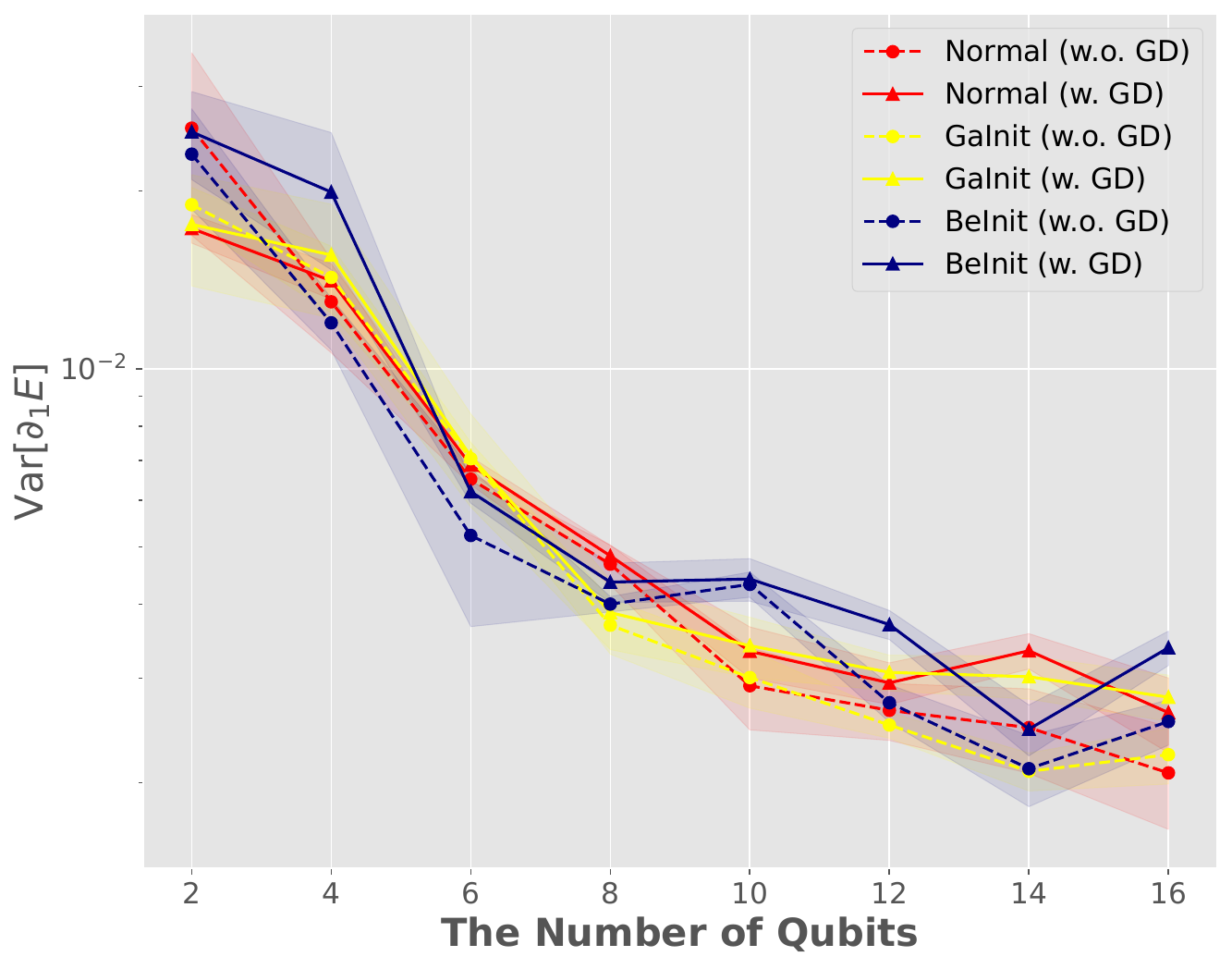}
    \includegraphics[width=0.49\textwidth]{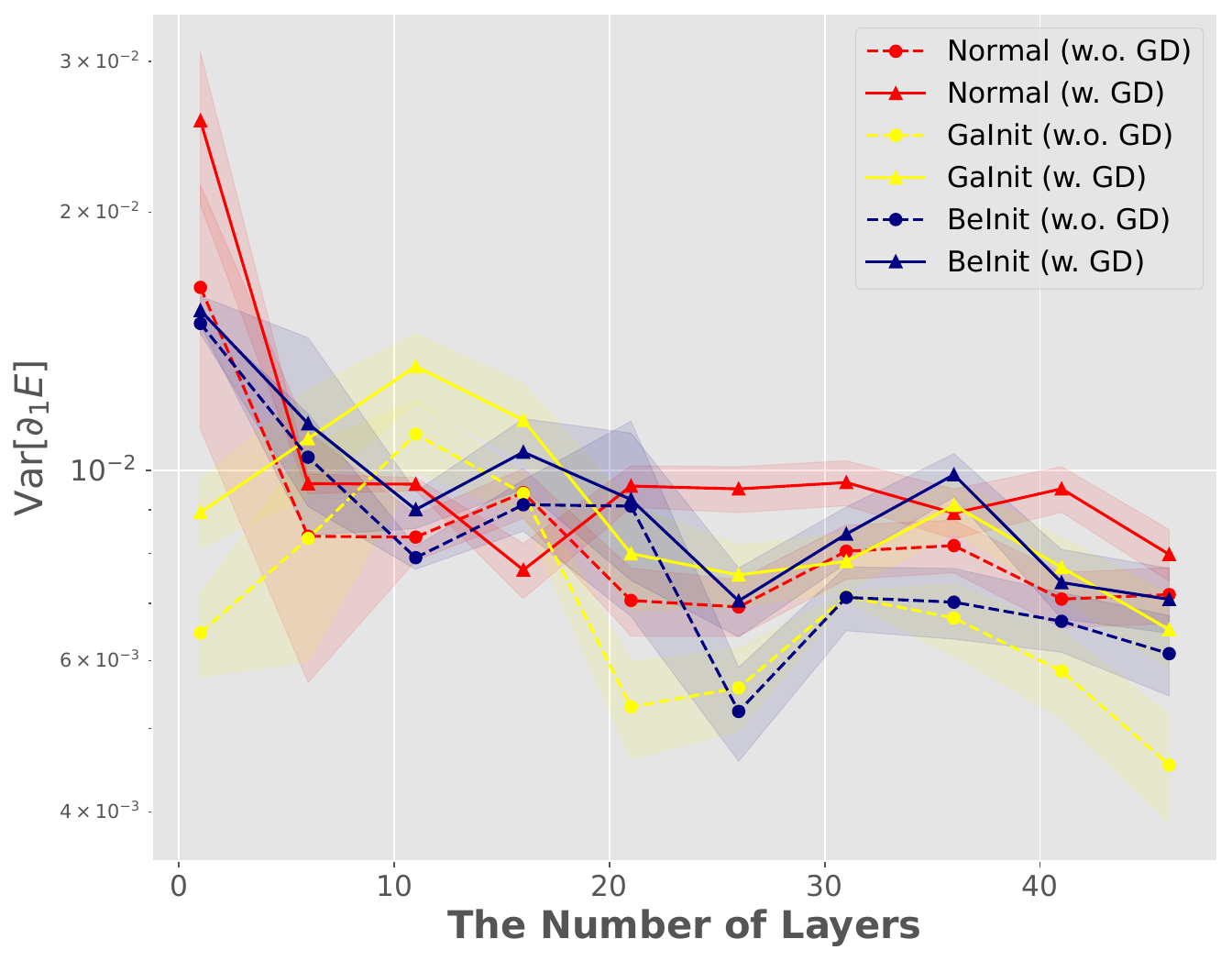}
    \caption{Iris}
  \end{subfigure}
  %\hfill
  \begin{subfigure}{0.5\linewidth}
    \centering  % include the 3rd and 4th images
    \includegraphics[width=0.49\textwidth]{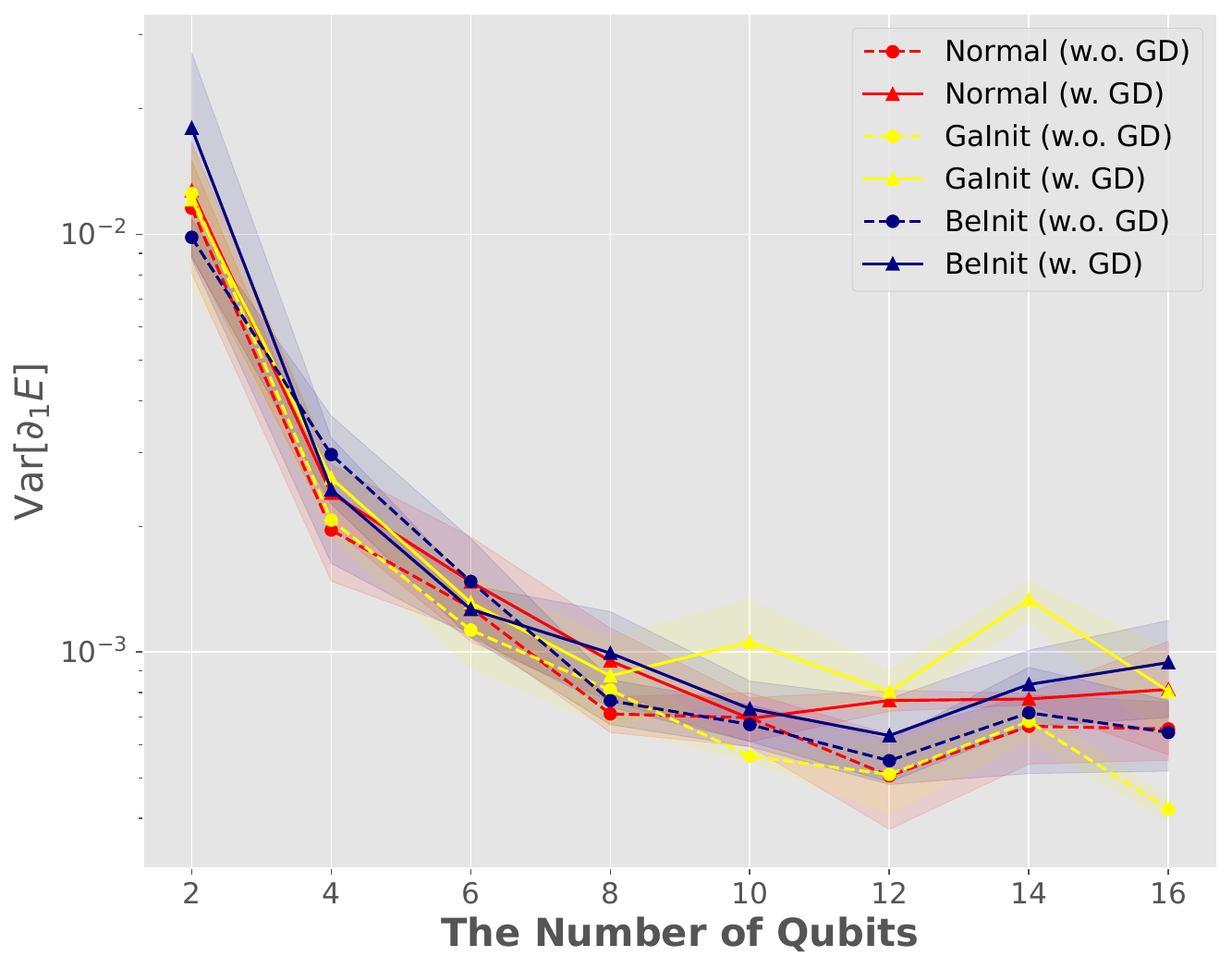}
    \includegraphics[width=0.49\textwidth]{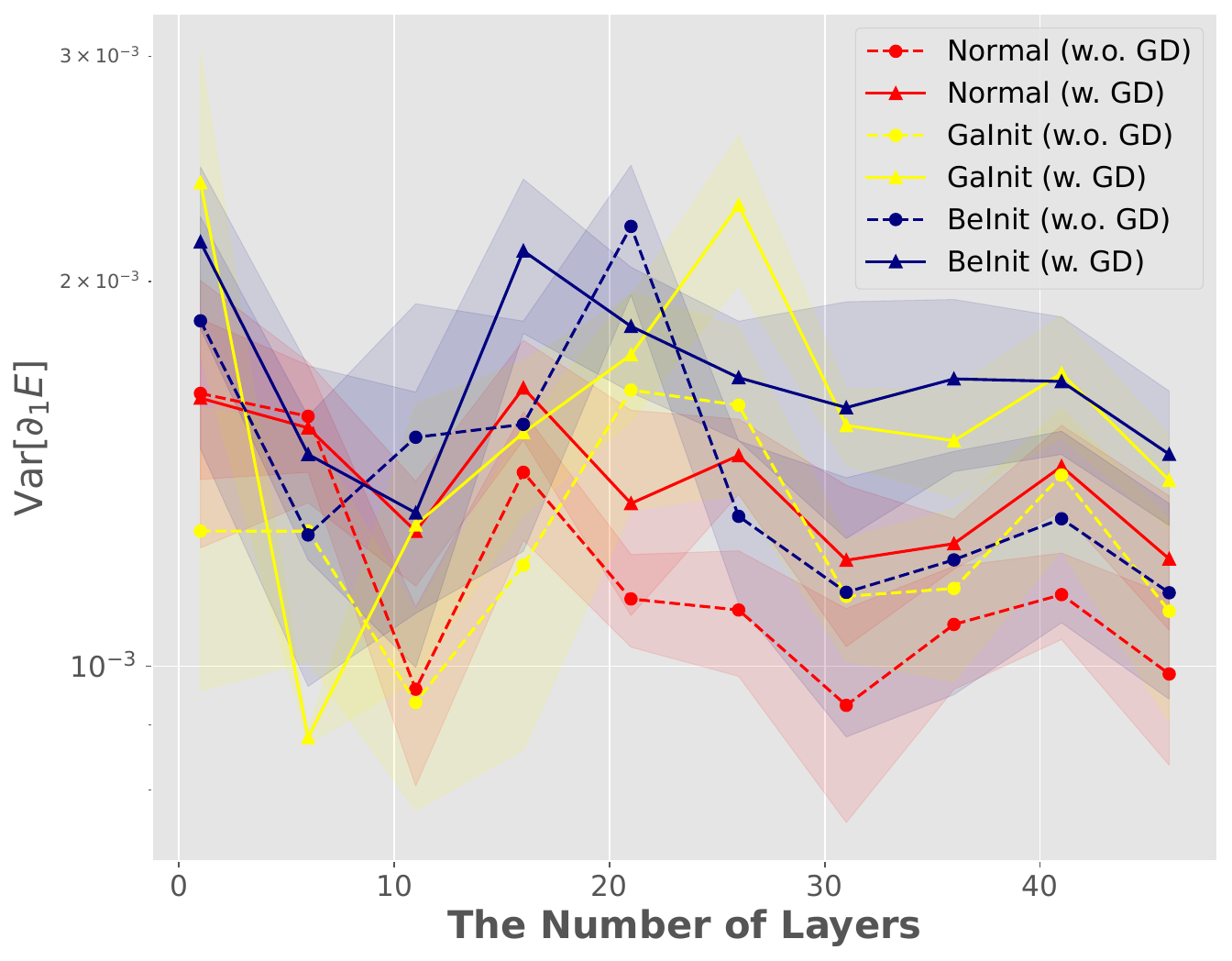}
    \caption{Wine}
  \end{subfigure}
  %\hfill
  \begin{subfigure}{0.5\linewidth}
    \centering  % include the 5th and 6th images
    \includegraphics[width=0.49\textwidth]{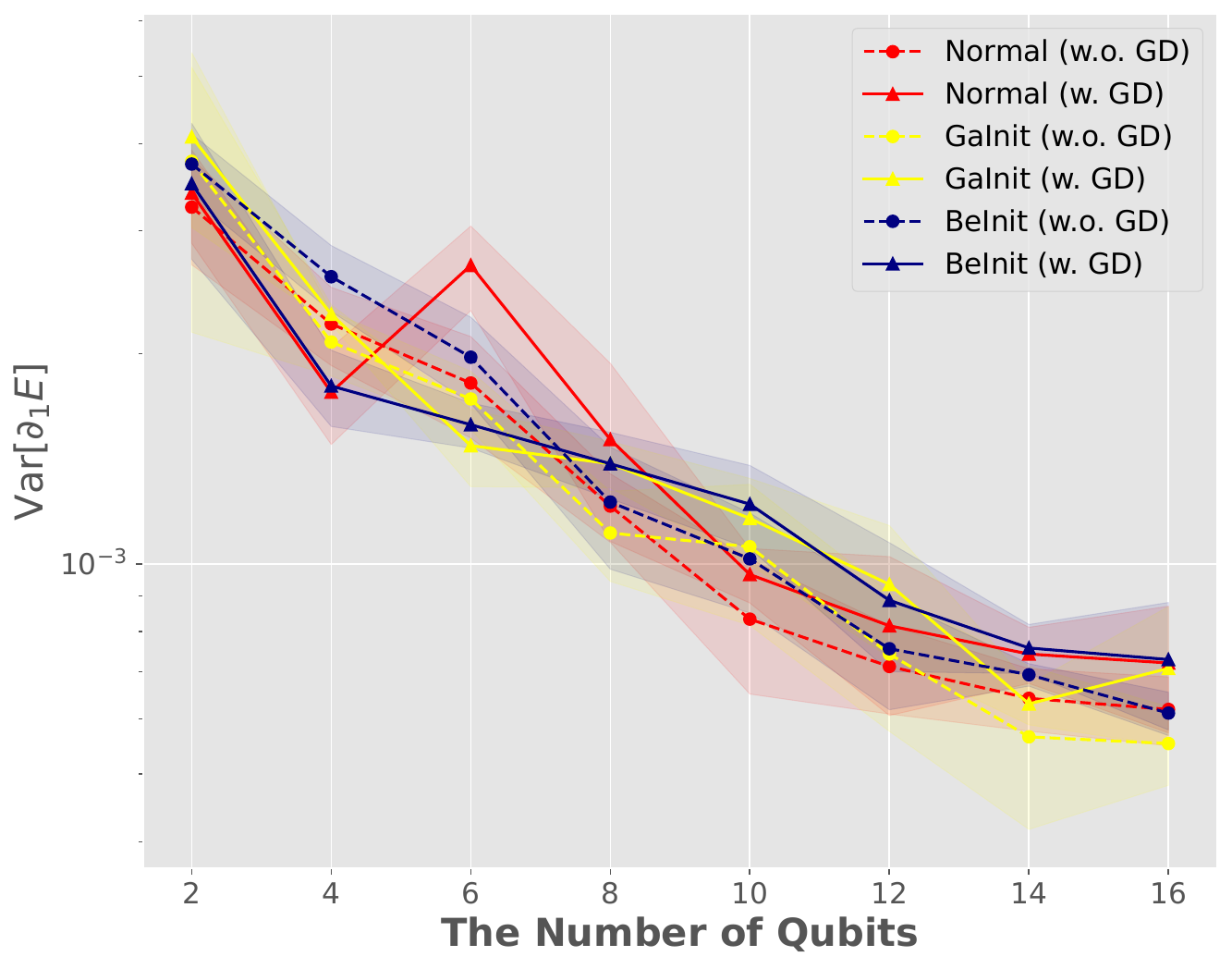}
    \includegraphics[width=0.49\textwidth]{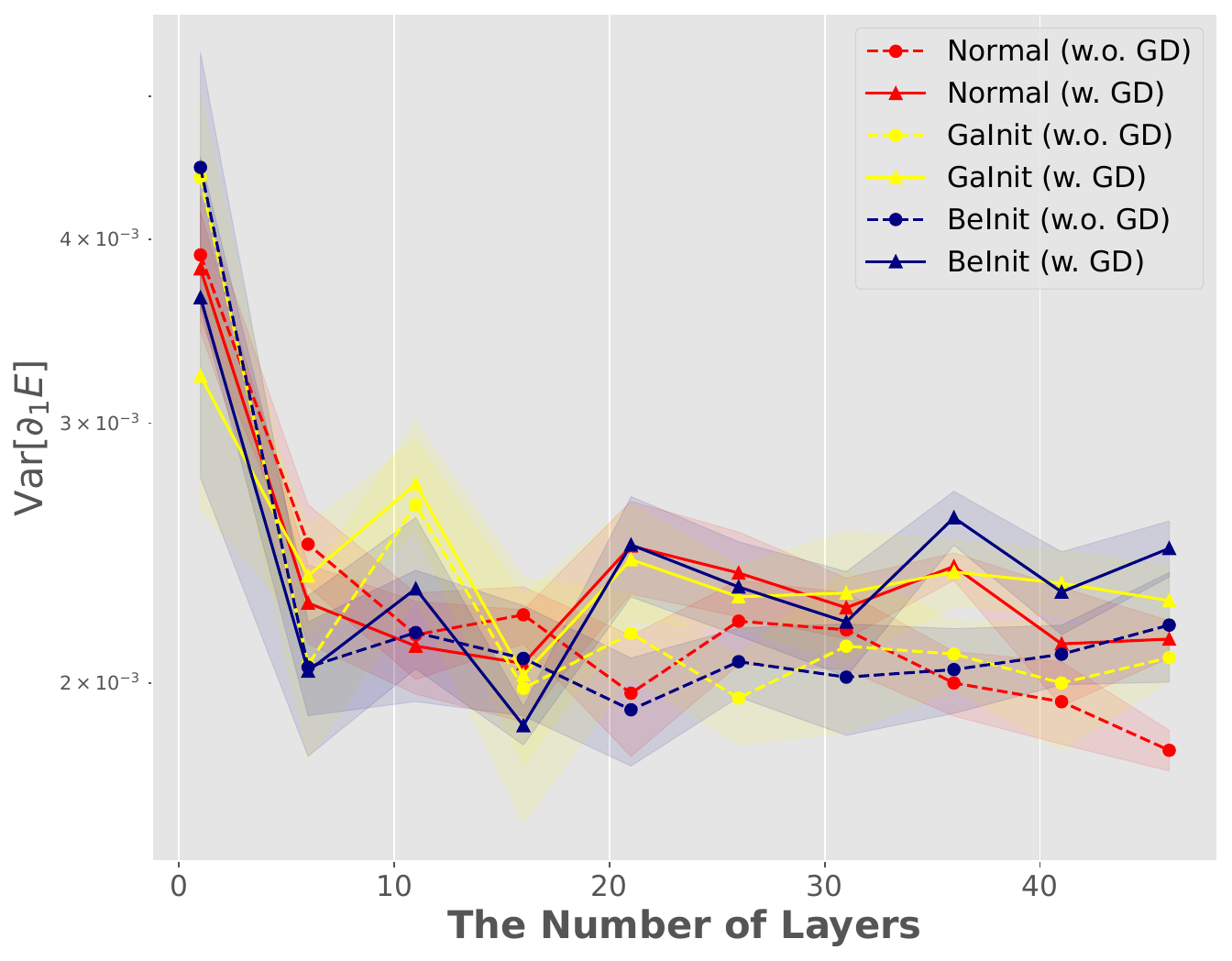}
    \caption{Titanic}
  \end{subfigure}
  %\hfill
  \begin{subfigure}{0.5\linewidth}
    \centering  % include the 7th and 8th images
    \includegraphics[width=0.49\textwidth]{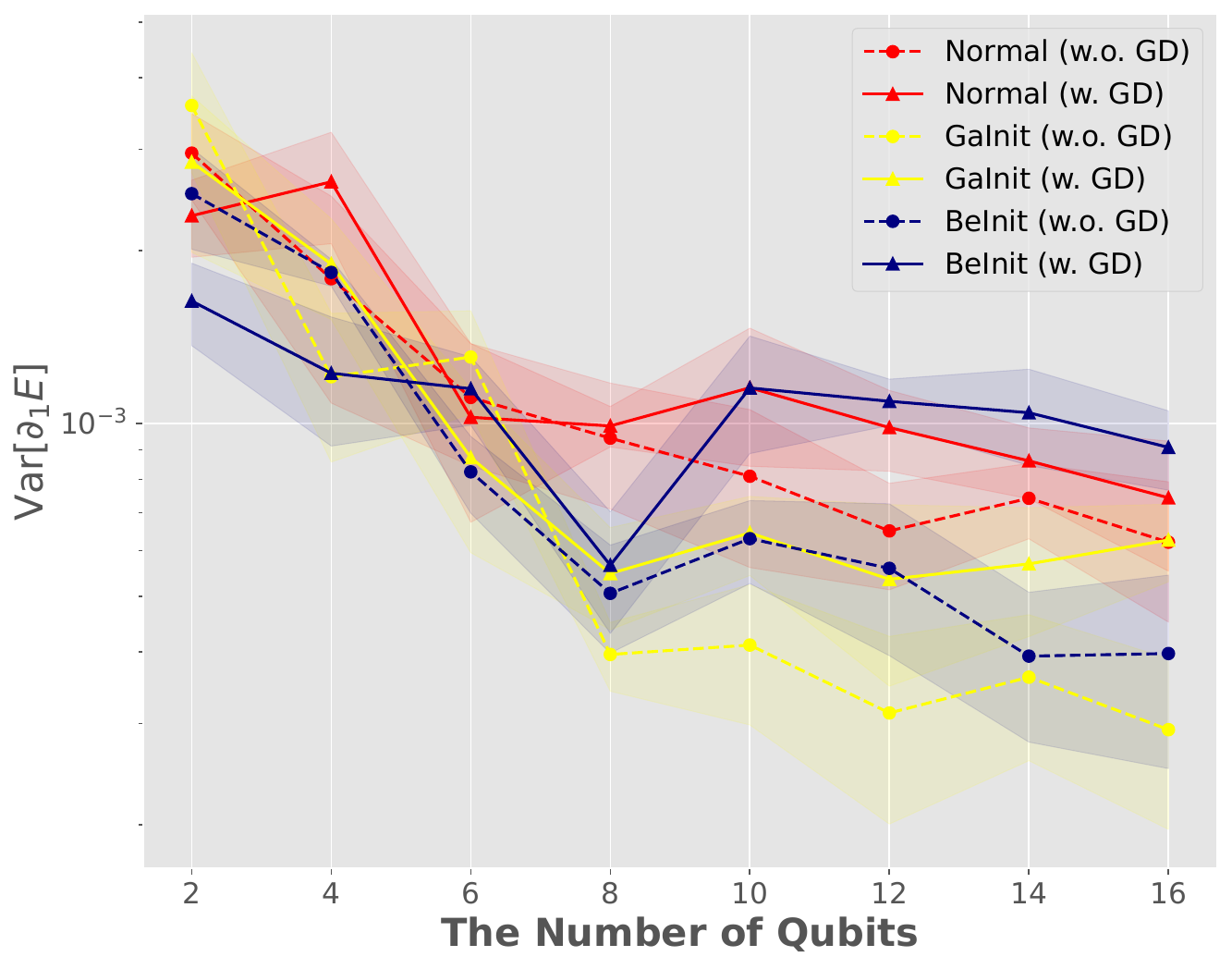}
    \includegraphics[width=0.49\textwidth]{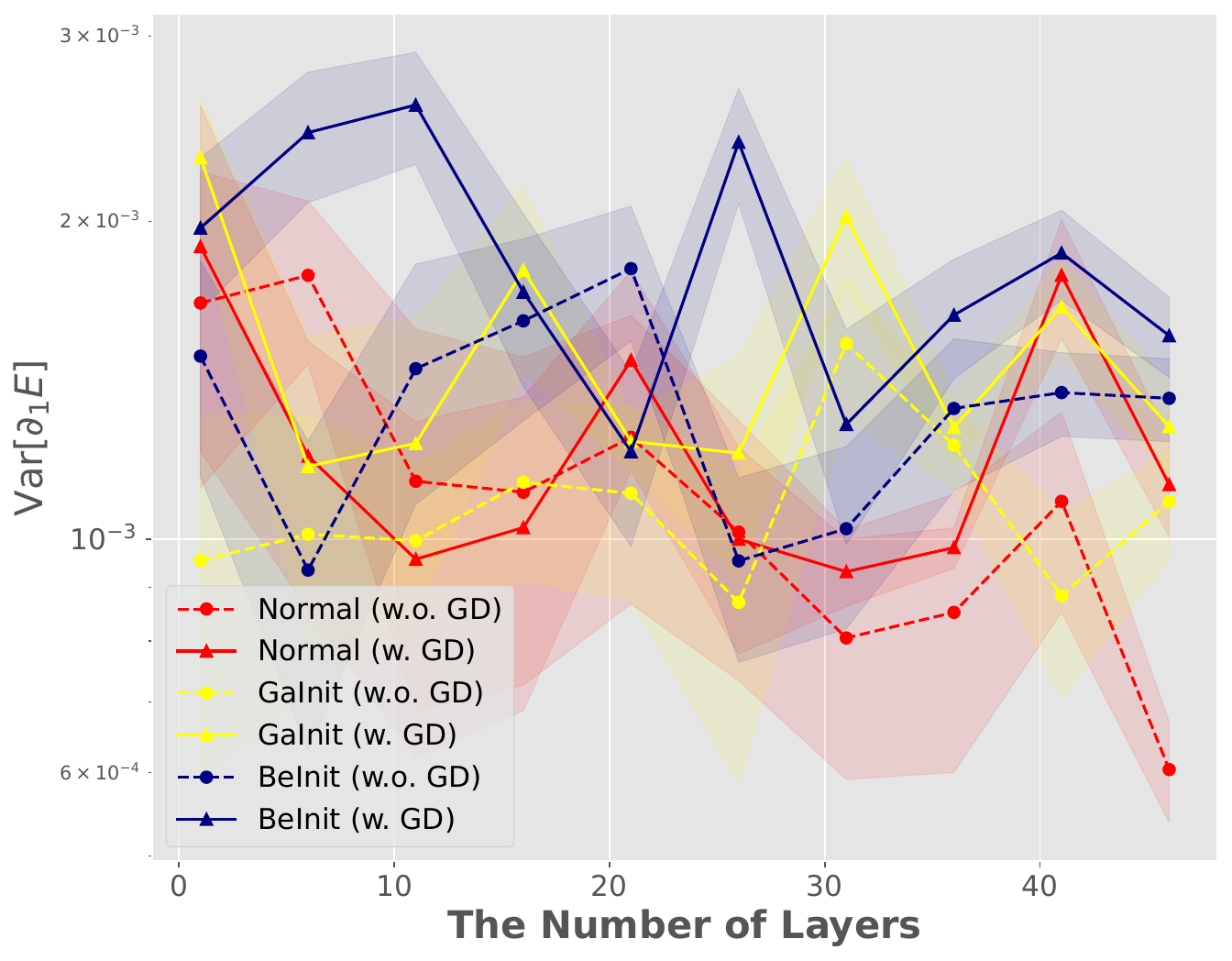}
    \caption{MNIST}
  \end{subfigure}
\caption{Investigation of the mechanism that regularizes model parameters by diffusing Gaussian noise along each iteration on five Gaussian-based methods. ``w. GD'' and ``w.o. GD'' denote whether or not we apply Gaussian noise diffusion.}
\label{fig:gd_exp}
\end{figure*}

\paragraph{Experimental Settings}
In the experiment, we evaluate our proposed method across four public datasets. {\bf Iris} is a classic machine-learning benchmark that measures various attributes of three-species iris flowers. {\bf Wine} is a well-known dataset that includes 13 attributes of chemical composition in wines. {\bf Titanic} contains historical data about passengers aboard the Titanic and is typically used to predict survival. {\bf MNIST} is a widely used small benchmark in computer vision. This benchmark consists of 28$\times$28 gray-scale images of handwritten digits from 0 to 9.

\begin{table}[h] % Dataset
\centering
%\footnotesize
%\scriptsize
%\setlength{\tabcolsep}{3.8pt}
\caption{Statistics of datasets. $\left| D \right|$, $\left| F \right|$, and $\left| C \right|$ denote the original number of instances, features, and classes, respectively. ``Splits'' denotes the split instances for the train, validation, and test data.}
\label{tab:data}
\begin{tabular}{ccccc}
  \toprule
    \textbf{Dataset} & {$\left| D \right|$} & {$\left| F \right|$} & {$\left| C \right|$} & {Splits} \\
    \midrule
    {\bf Iris} & 150 & 4 & 3 & 60:20:20 \\ %(60%:20%:20%)
    {\bf Wine} & 178 & 13 & 3 & 80:20:30 \\ %(62%:15%:23%)
    {\bf Titanic} & 891 & 11 & 2 & 320:80:179 \\ %(55%:14%:31%)
    {\bf MNIST} & 60,000 & 784 & 10 & 320:80:400 \\ %(40%:10%:50%)
  \bottomrule
\end{tabular}
\end{table}

We refer to the settings of BeInit~\cite{kulshrestha2022beinit} and examine the VQCs in binary classification, i.e., we sub-sample instances from the first two classes in each dataset to build a new subset. After sub-sampling, we re-scale the feature size no larger than the number of qubits. The statistics of original datasets and the data splits for train, validation, and test sets are provided in Table~\ref{tab:data}. Notably, the number of total sub-sampled instances is the sum of the split data. For example, in the Iris dataset, the number of sub-sampled instances is 100.

During training, we employ the Adam optimizer~\cite{kingma2014adam} to train VQCs with a learning rate of $1\times10^{-2}$ and a batch size of 20. The Optimization converged within 50 training epochs.
To assess the effectiveness of our proposed mechanisms, we ablatively apply the proposed mechanisms to baseline distributions, such as a Gaussian initial distribution, and then observe the gap between the two curves of gradient variance (whether or not our mechanism is applied to the baselines). We follow~\cite{McClean2018landscapes} to use gradient variance as an evaluation metric. Higher variance indicates better resistance to the gradient issues. We expect that after applying our mechanisms, the gap will become larger as the model size increases.
Based on the above settings, we aim to ablatively investigate whether our proposed mechanisms can facilitate the trainability of VQCs in the following subsections.

\paragraph{Regularization with Prior Knowledge of the Training Data Can Help Alleviate Barren Plateaus}
We conduct experiments to investigate whether prior knowledge can contribute to initializing the model parameters along different qubits or layers.
In experiments, we include three groups of initial distributions. For each group, we examine two scenarios, applying prior distributions to the initial distributions (``w. Pr'') or not (``w.o. Pr'').
As presented in Figure~\ref{fig:pr_exp}, we repeat experiments five times and plot curves of the first-layer variance for three classic initialization methods (the rest of the five methods are presented in Figure~\ref{fig:pr_exp2}). We observe that in most cases, the variance in the first layer will gradually decrease as the number of qubits or layers increases. Besides, we expect that the solid lines are higher than the dashed lines along with different qubits or layers, which demonstrates that incorporating the prior distribution of the train data in initialization can maintain higher variance, thereby mitigating barren plateau issues.

\paragraph{Regularization with Gaussian Noise Diffusion Can Help Avoid Being Trapped in Saddle Points}
We conduct another ablation study to examine the effectiveness of our proposed regularization strategy. Specifically, we apply our proposed regularization strategy to both the Normal distribution and two state-of-the-art methods, Gaussian initialization (GaInit) and Beta initialization (BeInit), and examine whether diffusing Gaussian noise on the model parameters along each training epoch as a regularization can help avoid being trapped in saddle points. We expect that this mechanism can increase volatility while alleviating the degradation of gradient variance during training.
As presented in Figure~\ref{fig:gd_exp}, we repeat experiments five times and plot curves of the first-layer variance for each method. We observe that in most cases, the solid lines (``w. GD'') are higher than the dashed lines (``w.o. GD''). The results indicate that after applying Gaussian noise diffusion to the model parameters, the volatility of gradient variance stays higher so the optimization has a higher probability of avoiding being trapped in saddle points, whereas the gradient variance decreases much slower (i.e., the gap between two scenarios, ``w.o. GD'' and ``w. GD'', becomes wider) as the number of qubits or layers increases, verifying the effectiveness of our proposed mechanism.

\begin{figure*}[h]
  %\hfill
  \begin{subfigure}{0.5\linewidth}
    \centering  % include the 1st and 2nd images
    \includegraphics[width=0.49\textwidth]{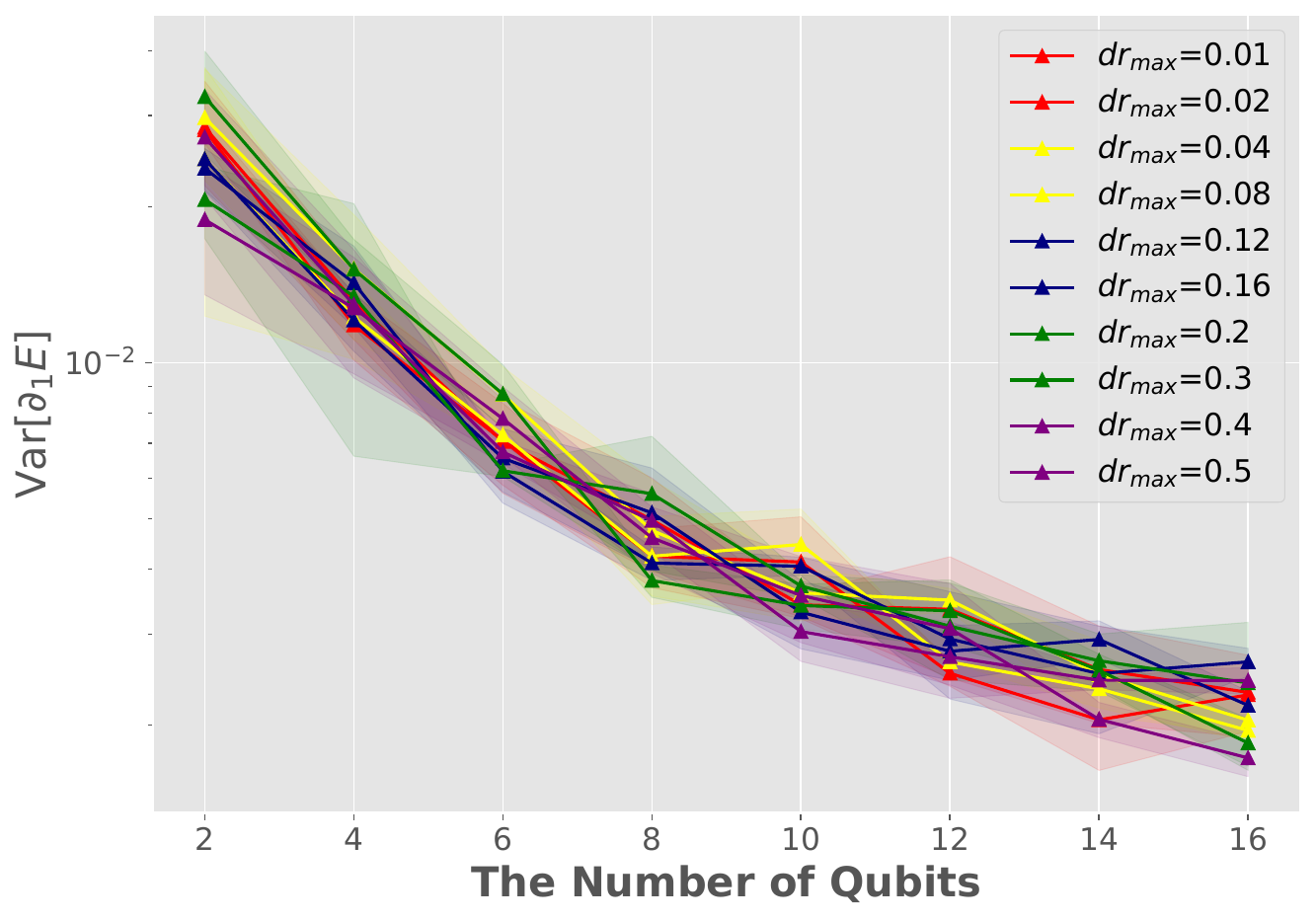}
    \includegraphics[width=0.49\textwidth]{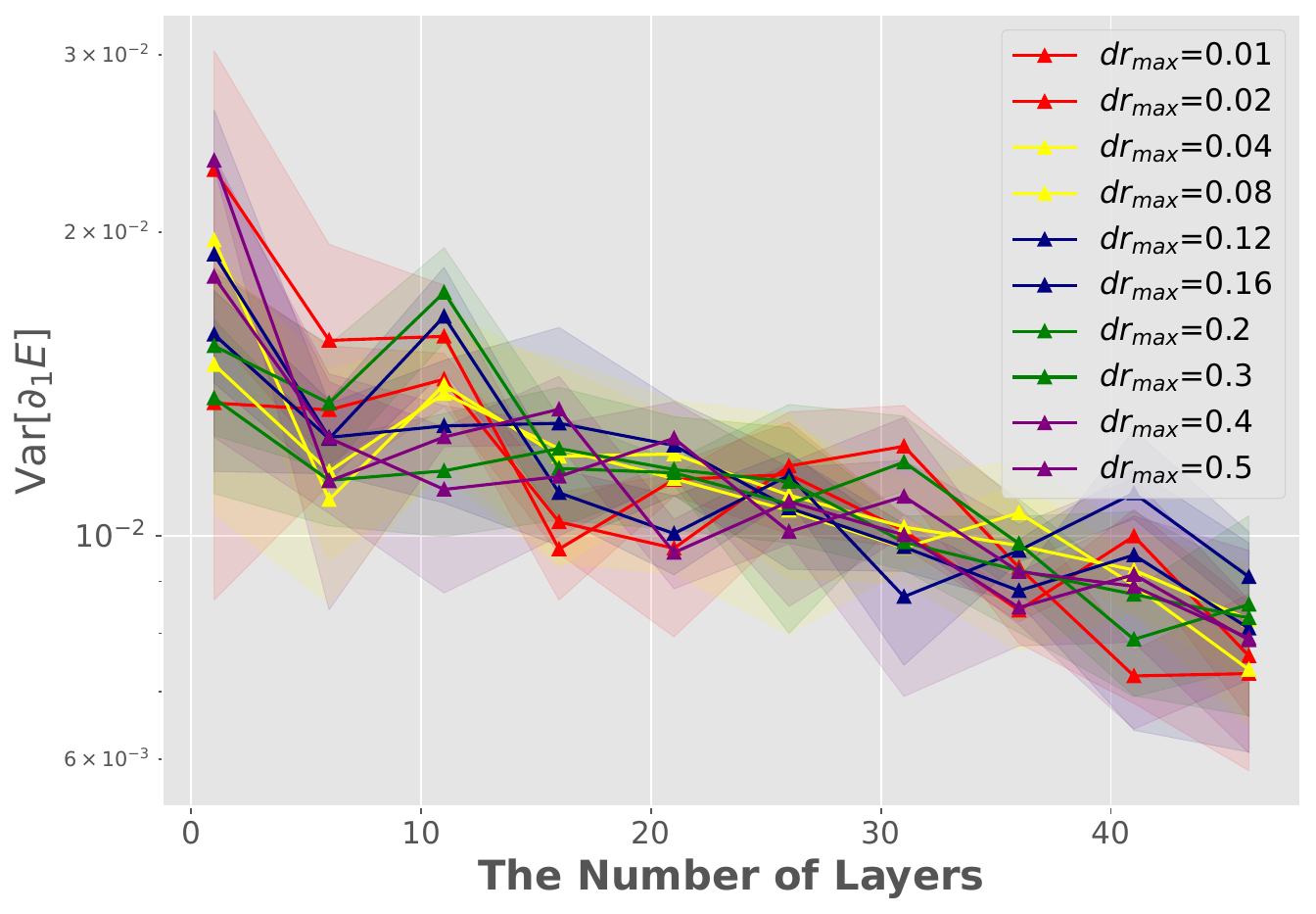}
    \caption{Iris}
  \end{subfigure}
  %\hfill
  \begin{subfigure}{0.5\linewidth}
    \centering  % include the 3rd and 4th images
    \includegraphics[width=0.49\textwidth]{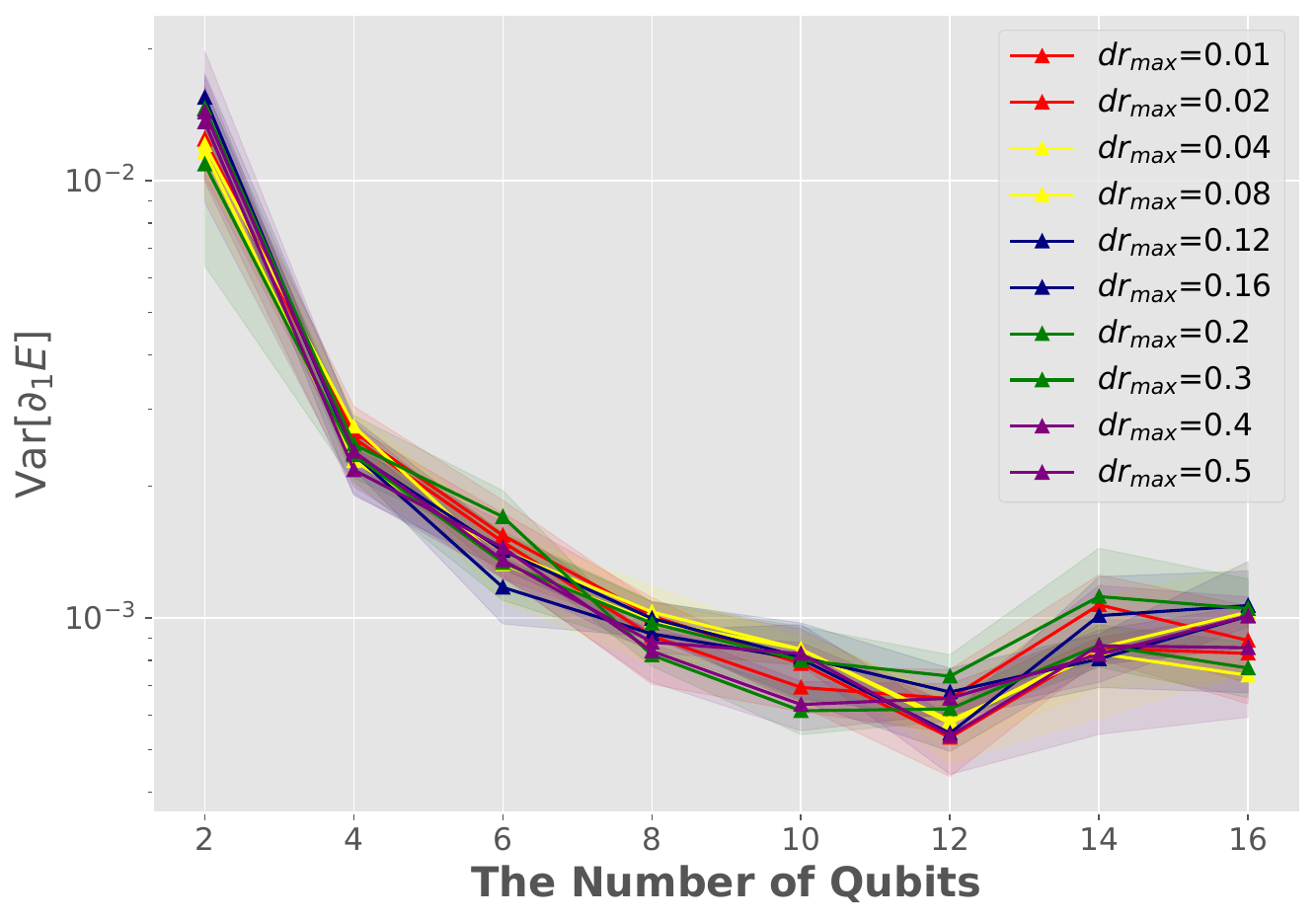}
    \includegraphics[width=0.49\textwidth]{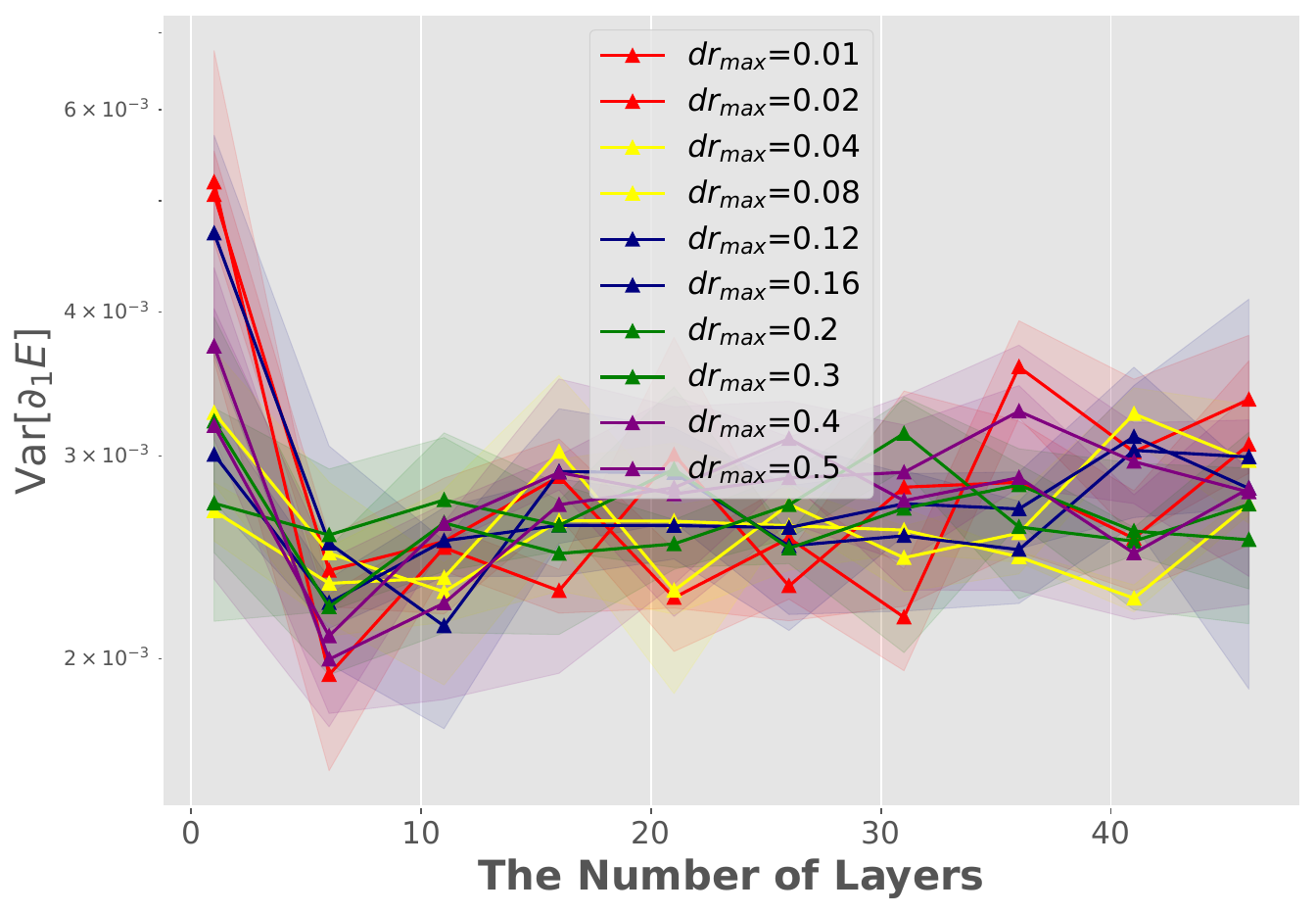}
    \caption{Wine}
  \end{subfigure}
  %\hfill
  \begin{subfigure}{0.5\linewidth}
    \centering  % include the 5th and 6th images
    \includegraphics[width=0.49\textwidth]{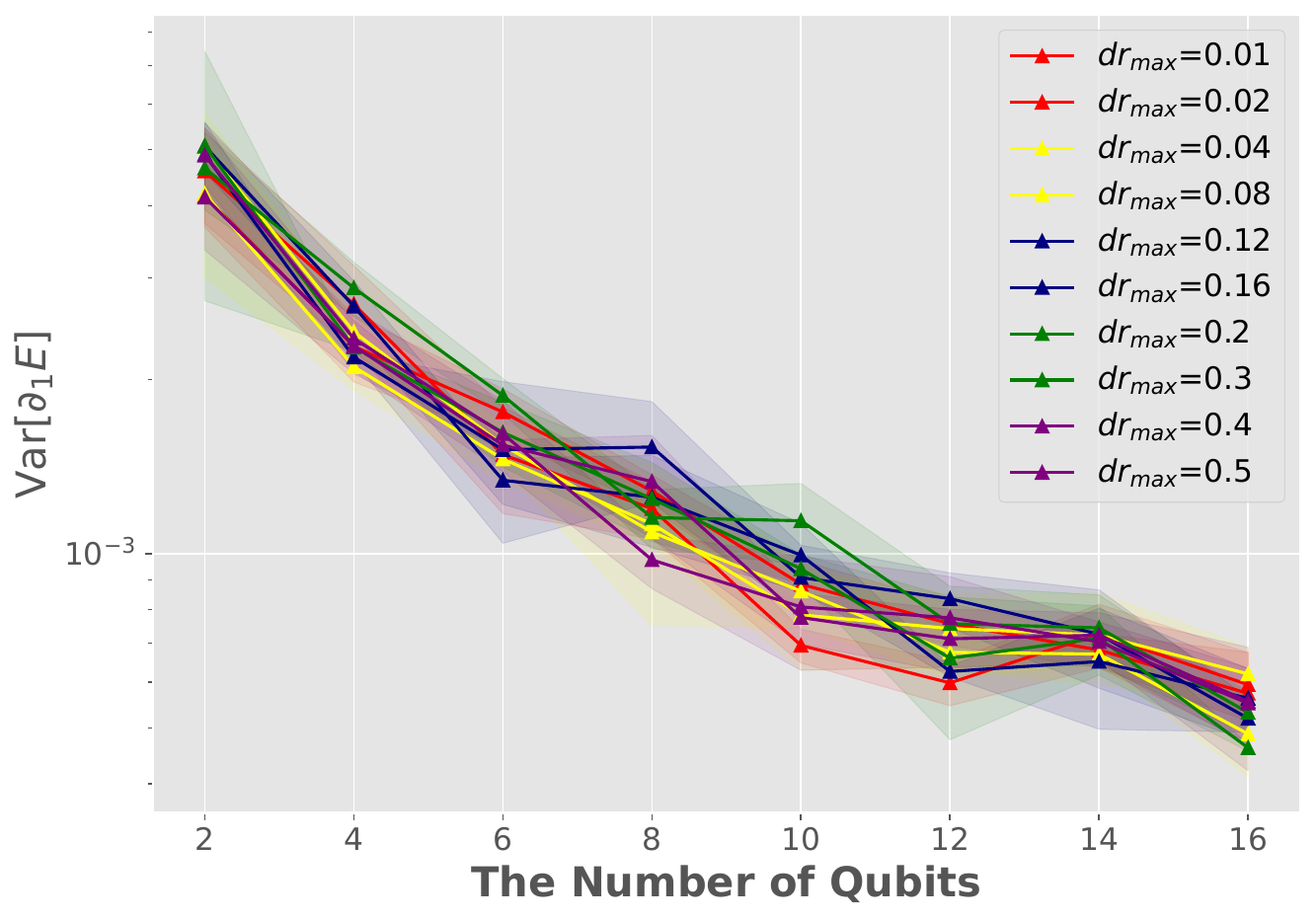}
    \includegraphics[width=0.49\textwidth]{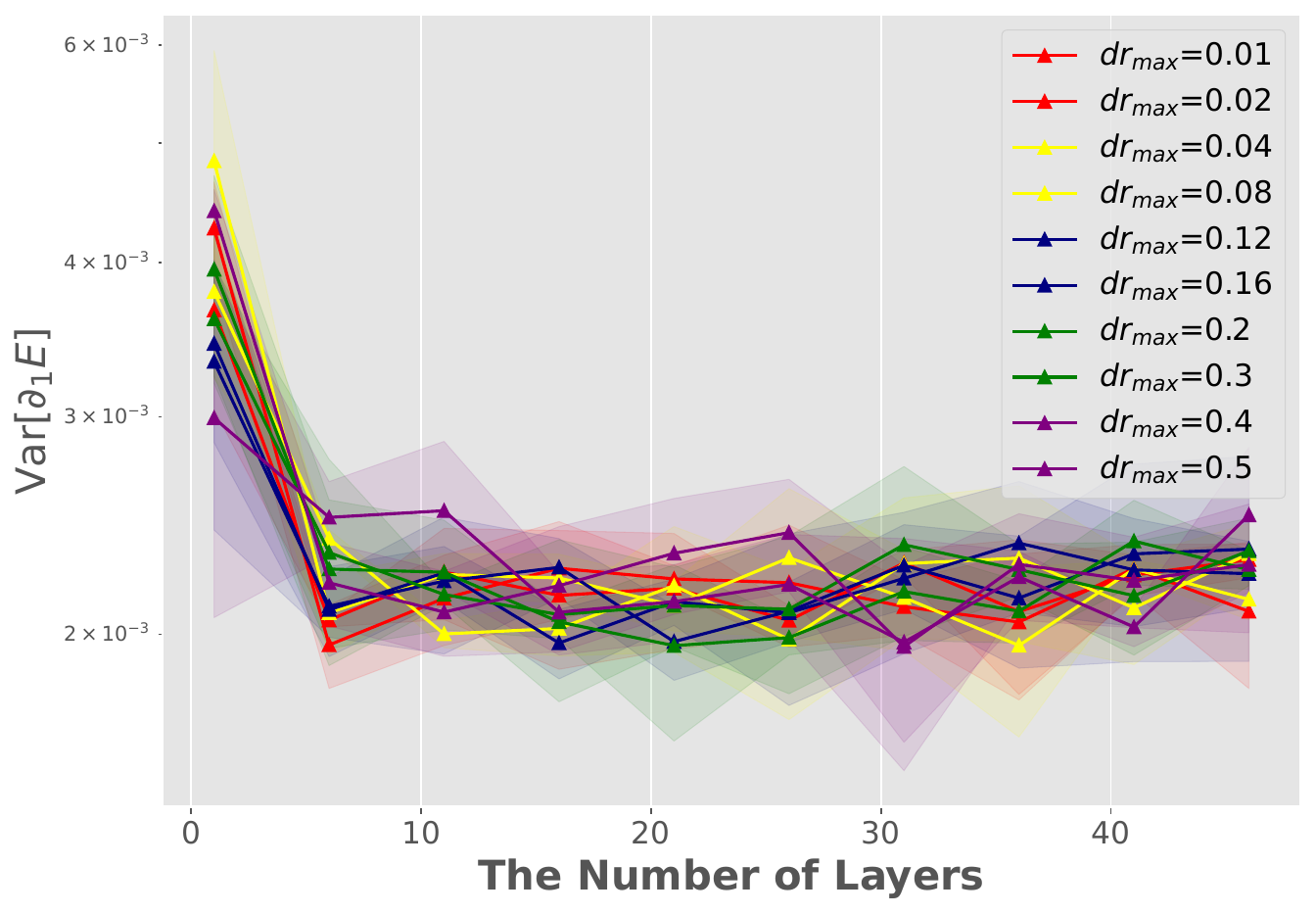}
    \caption{Titanic}
  \end{subfigure}
  %\hfill
  \begin{subfigure}{0.5\linewidth}
    \centering  % include the 7th and 8th images
    \includegraphics[width=0.49\textwidth]{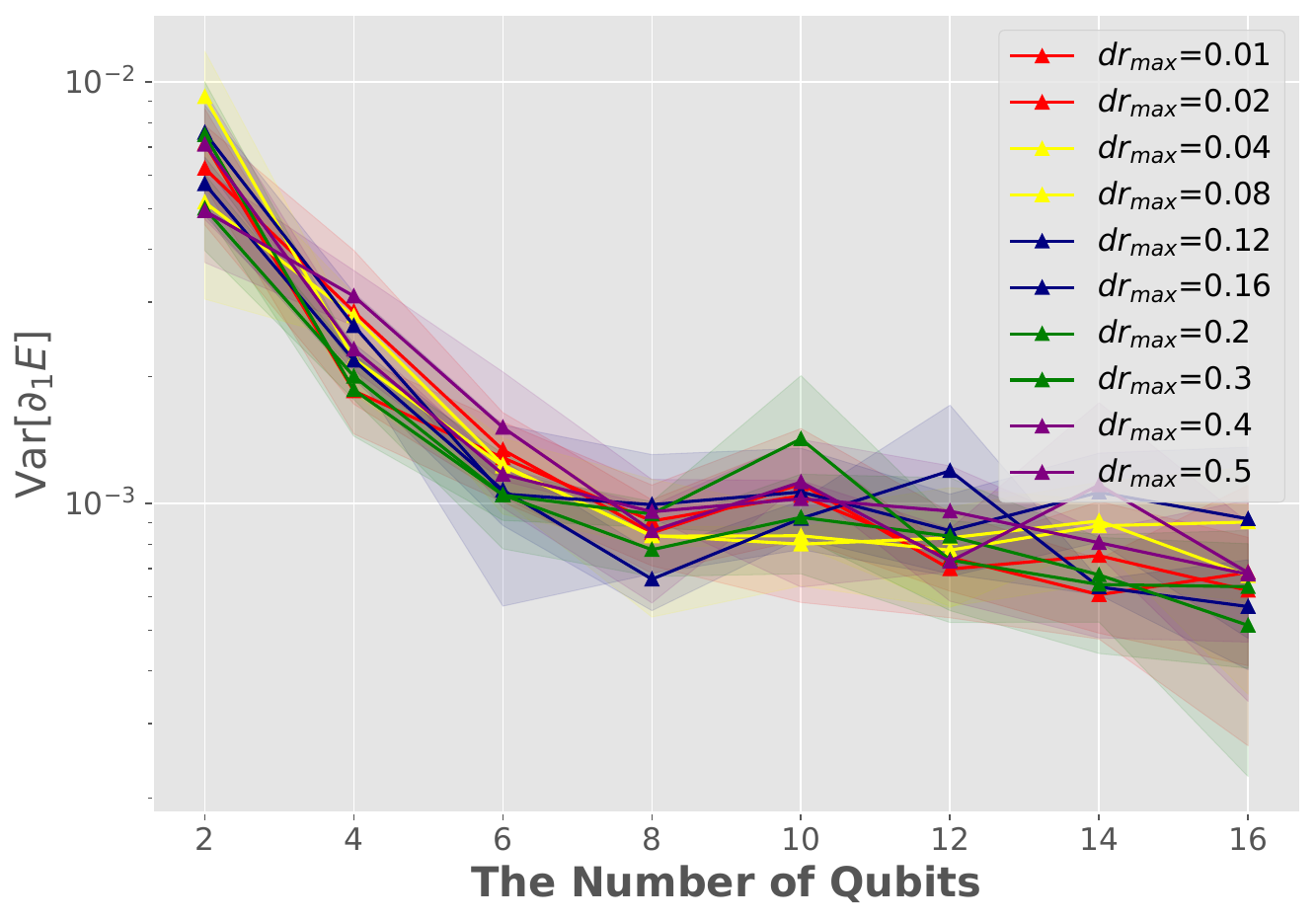}
    \includegraphics[width=0.49\textwidth]{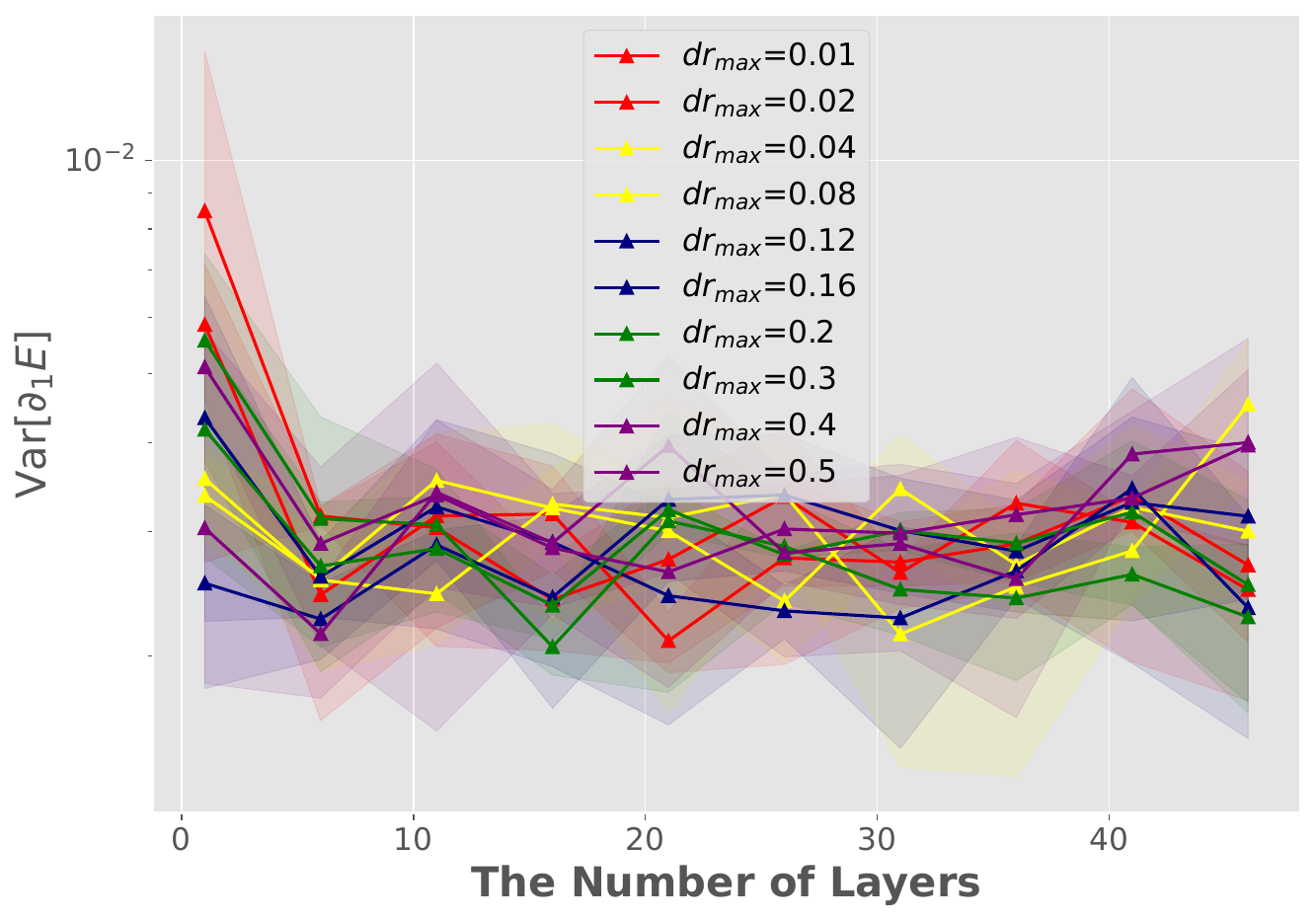}
    \caption{MNIST}
  \end{subfigure}
\caption{Analysis of the hyperparameter, $dr_{max}$, along different numbers of qubits or layers on four public datasets.}
\label{fig:dr_exp}
\end{figure*}

%\begin{wraptable}{R}{0.2\textwidth}
\begin{table}[h]
\centering
\caption{The optimal hyperparameter, max diffusion rate ($dr_{max}$) in each scenario. We report the results on Normal distribution as an example.}
\label{tab:drmax}
\begin{tabular}{ccc}
  \toprule
    {Dataset} & {Scenario} & {\bf $dr_{max}$} \\
    \midrule
    \multirow{2}{*}{Iris} & Qubits & 0.30 \\
                          & Layers & 0.02 \\
    \hline
    \multirow{2}{*}{Wine} & Qubits & 0.16 \\
                          & Layers & 0.01 \\
    \hline
    \multirow{2}{*}{Titanic} & Qubits & 0.20 \\
                             & Layers & 0.50 \\
    \hline
    \multirow{2}{*}{MNIST} & Qubits & 0.04 \\
                           & Layers & 0.02 \\
  \bottomrule
\end{tabular}
\end{table}
%\end{wraptable}

\paragraph{Analysis of Hyperparameter}
Besides verifying the effectiveness of our proposed methods, we fix the min diffusion rate ($dr_{min}$) as $1\times10^{-4}$ and analyze the sensitivity of the key hyperparameter, max diffusion rate ($dr_{max}$), along different qubits or layers on the validation set. In this experiment, we simultaneously consider both mechanisms, i.e., applying prior knowledge of the train data in initialization and further diffusing Gaussian noise on model parameters along each training epoch. We repeat experiments five times on the Normal distribution as an example and present curves of the first-layer variance for different $dr_{max}$ in Figure~\ref{fig:dr_exp}. We further compute the mean value of each curve (under different $dr_{max}$) and select the $dr_{max}$ in each scenario (either ``qubits'' or ``layers'' in each dataset) based on the maximum mean value. The optimal $dr_{max}$ for each scenario in this study is reported in Table~\ref{tab:drmax}.

%% file: 5appendix.tex
\section{Appendix}
\label{sec:appendix}
In the appendix, we first present additional experimental results to validate the generalization of our proposed mechanism. Besides, we present the baseline VQC and the settings for our hardware and software.

\begin{figure*}[h]
\centering
  %\hfill
  \begin{subfigure}{0.49\linewidth}
    \centering  % include the 1st and 2nd images
    \includegraphics[width=0.49\textwidth]{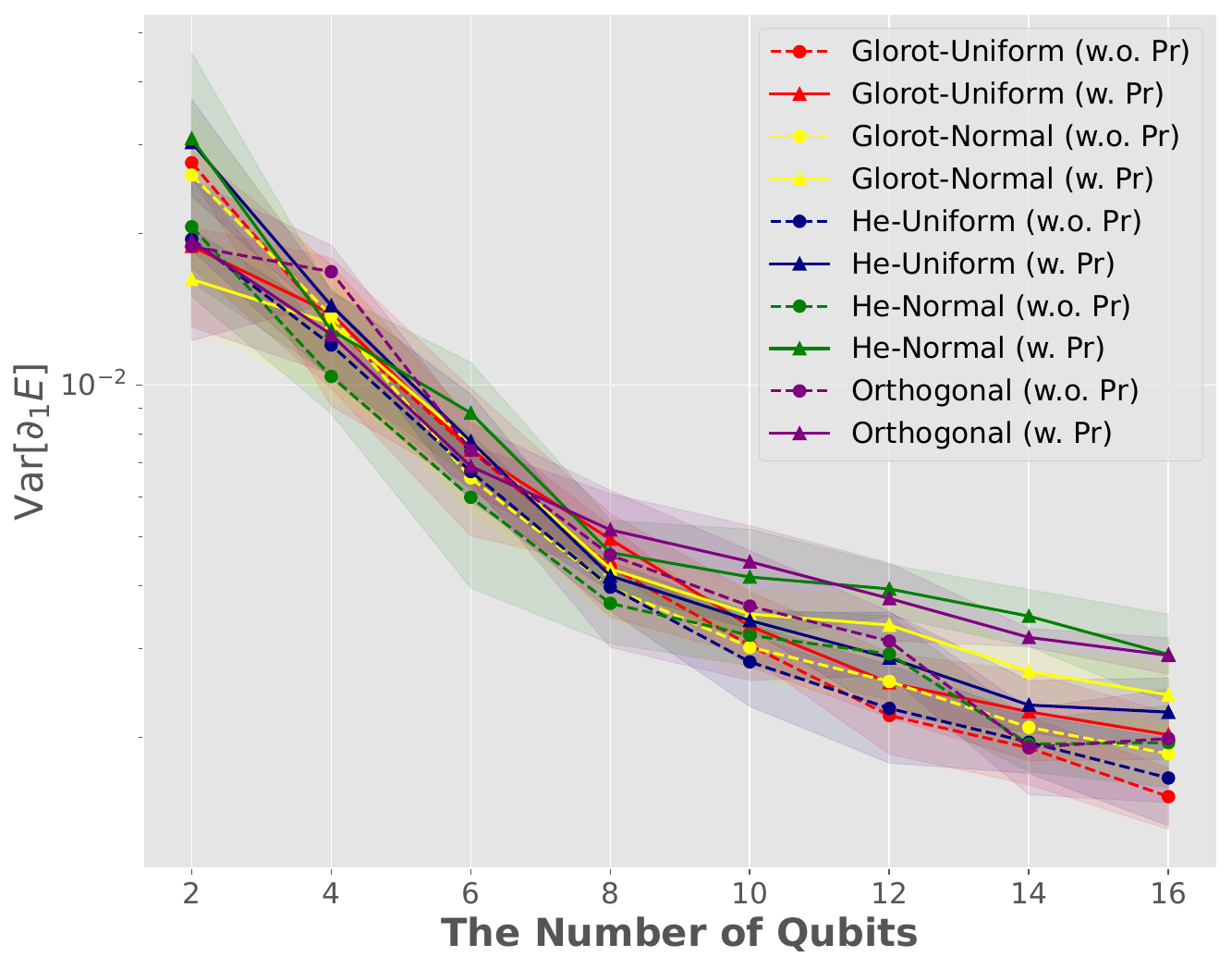}
    \includegraphics[width=0.49\textwidth]{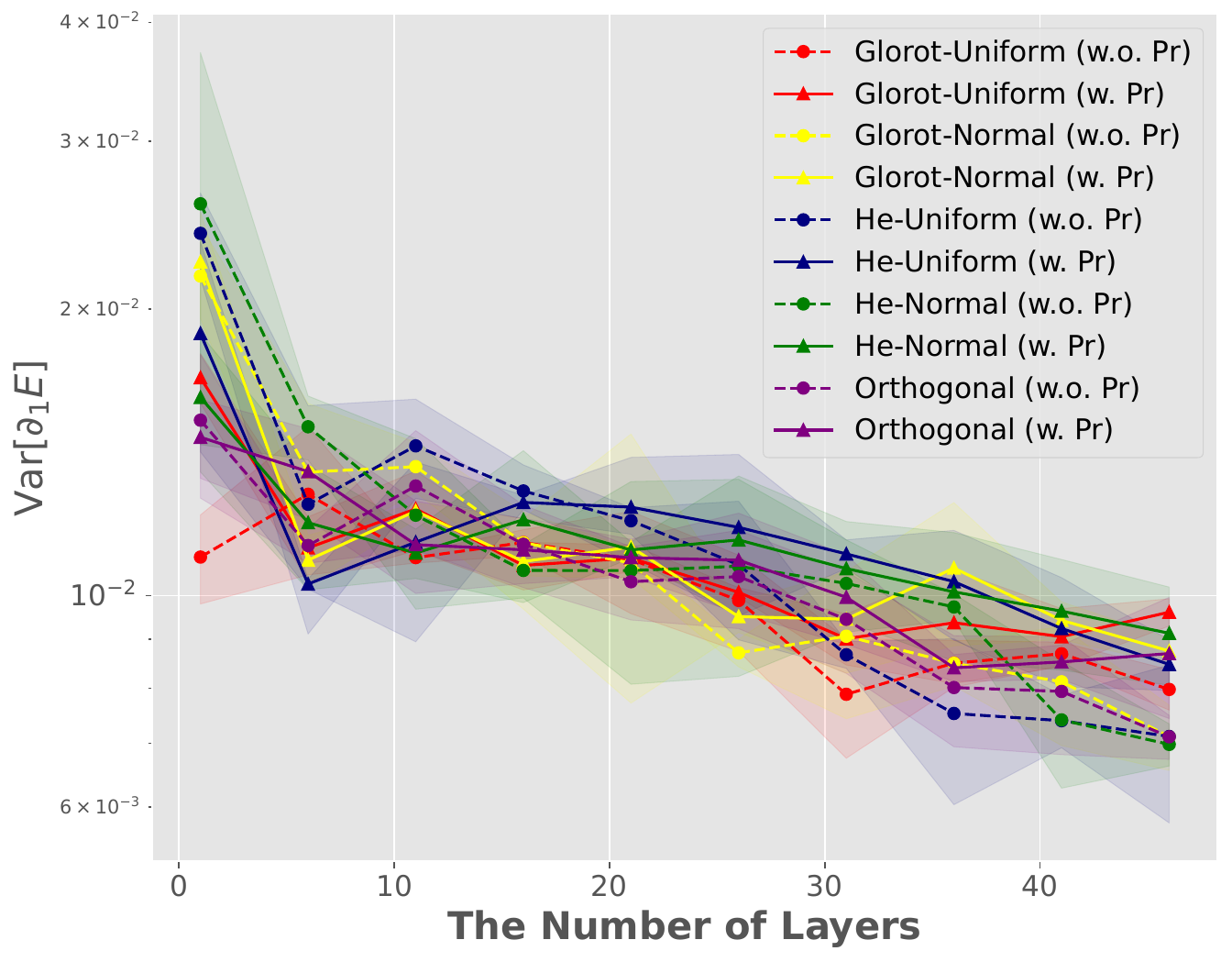}
    \caption{Iris}
  \end{subfigure}
  %\hfill
  \begin{subfigure}{0.49\linewidth}
    \centering  % include the 3rd and 4th images
    \includegraphics[width=0.49\textwidth]{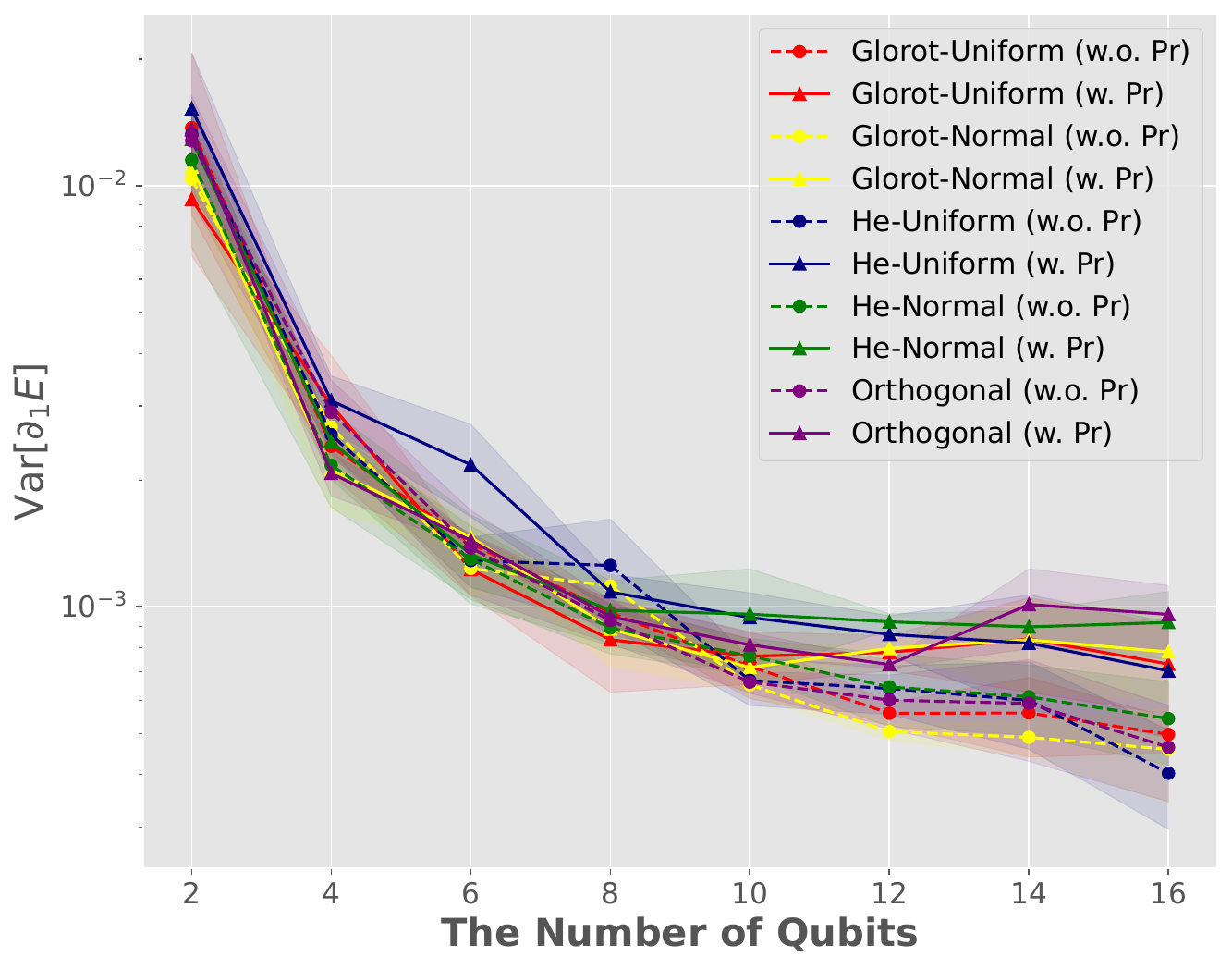}
    \includegraphics[width=0.49\textwidth]{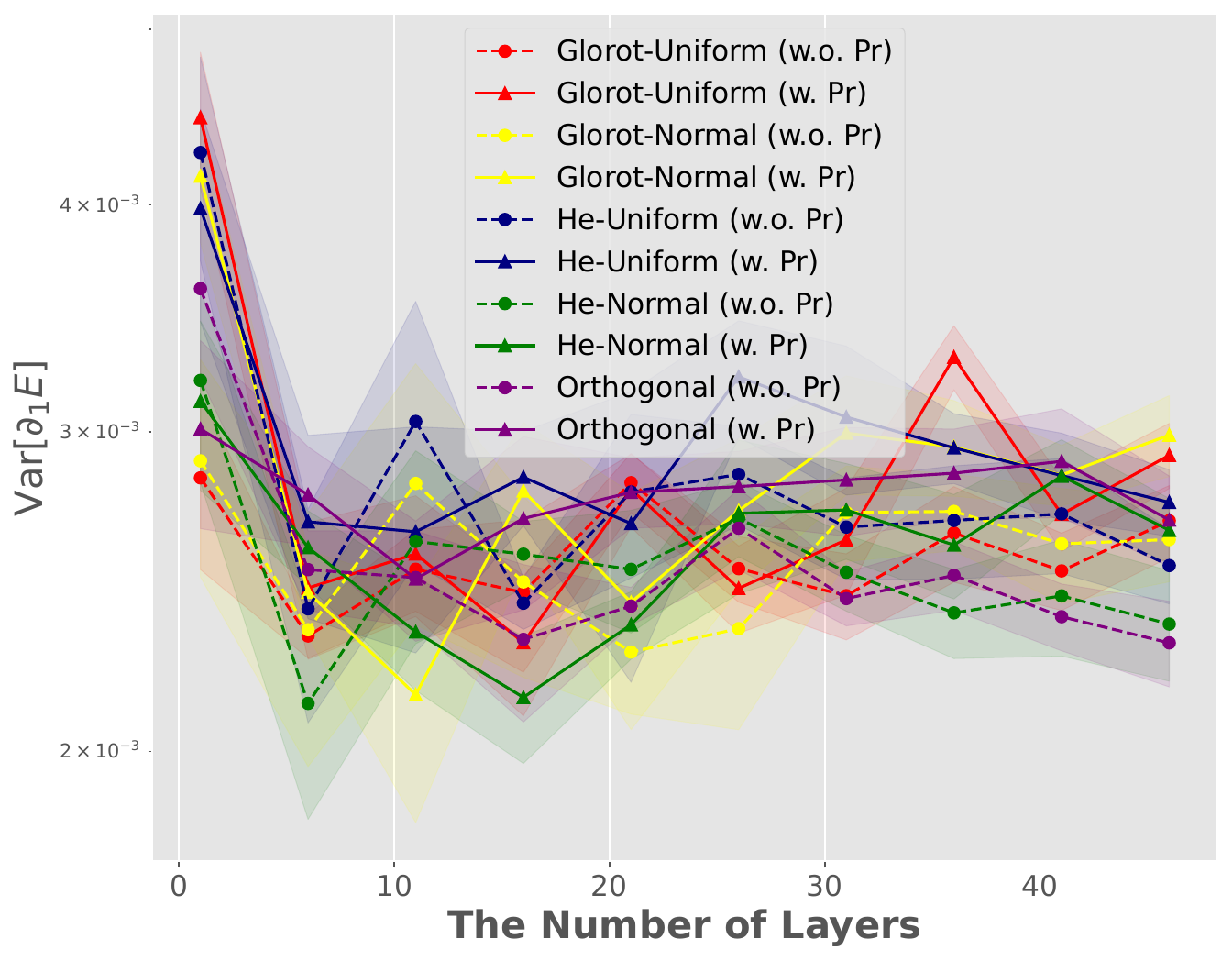}
    \caption{Wine}
  \end{subfigure}
  %\hfill
  \begin{subfigure}{0.49\linewidth}
    \centering  % include the 5th and 6th images
    \includegraphics[width=0.49\textwidth]{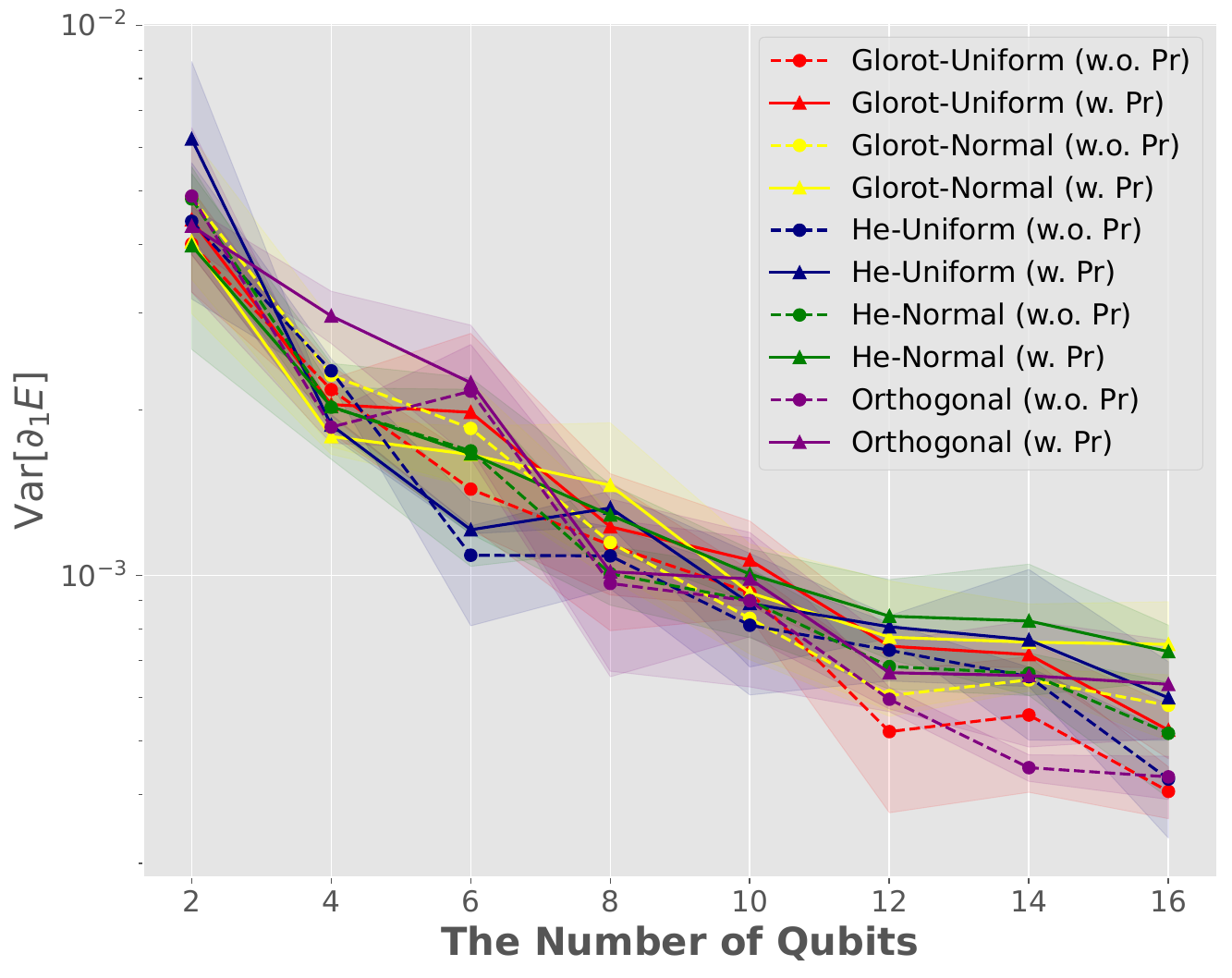}
    \includegraphics[width=0.49\textwidth]{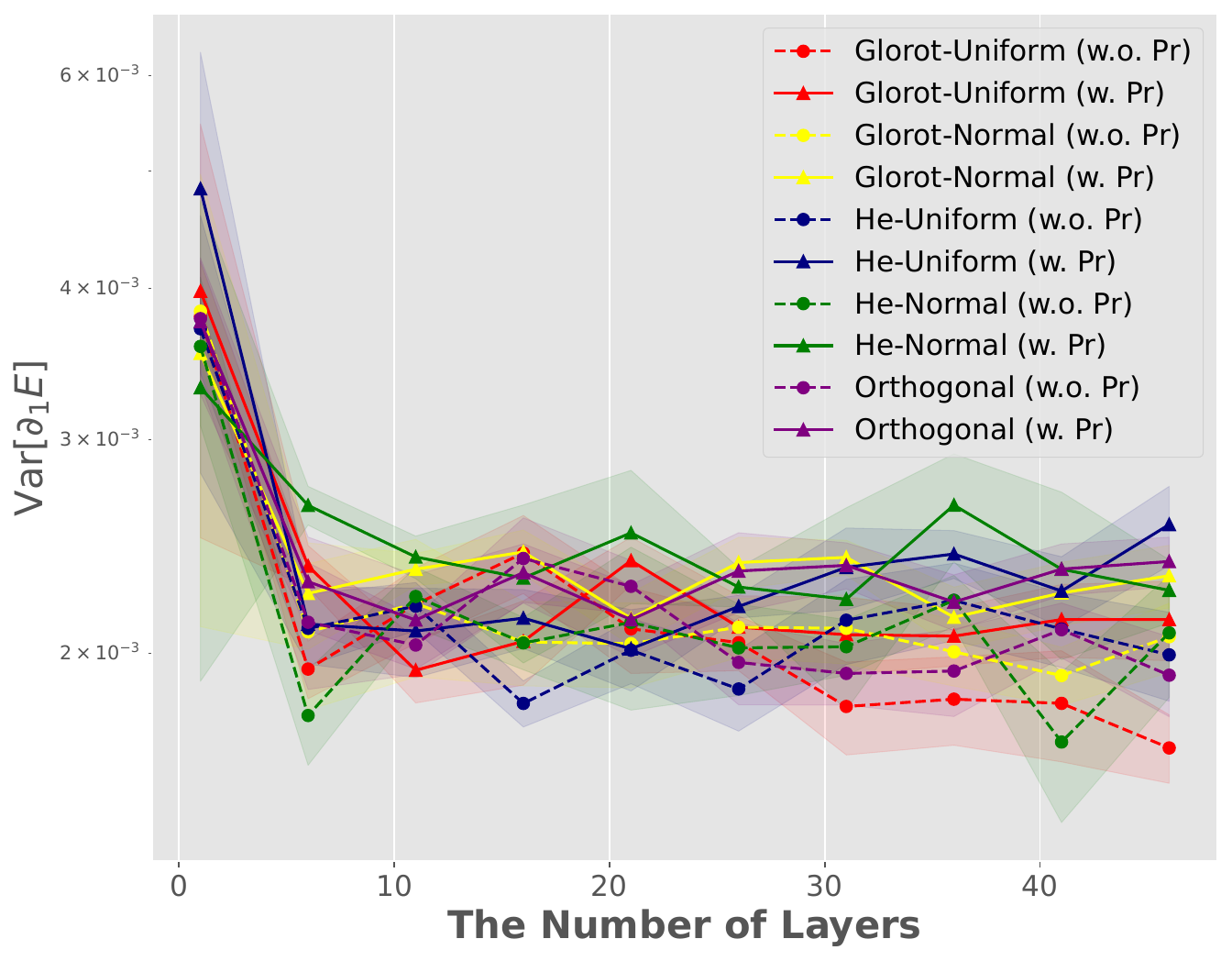}
    \caption{Titanic}
  \end{subfigure}
  %\hfill
  \begin{subfigure}{0.49\linewidth}
    \centering  % include the 7th and 8th images
    \includegraphics[width=0.49\textwidth]{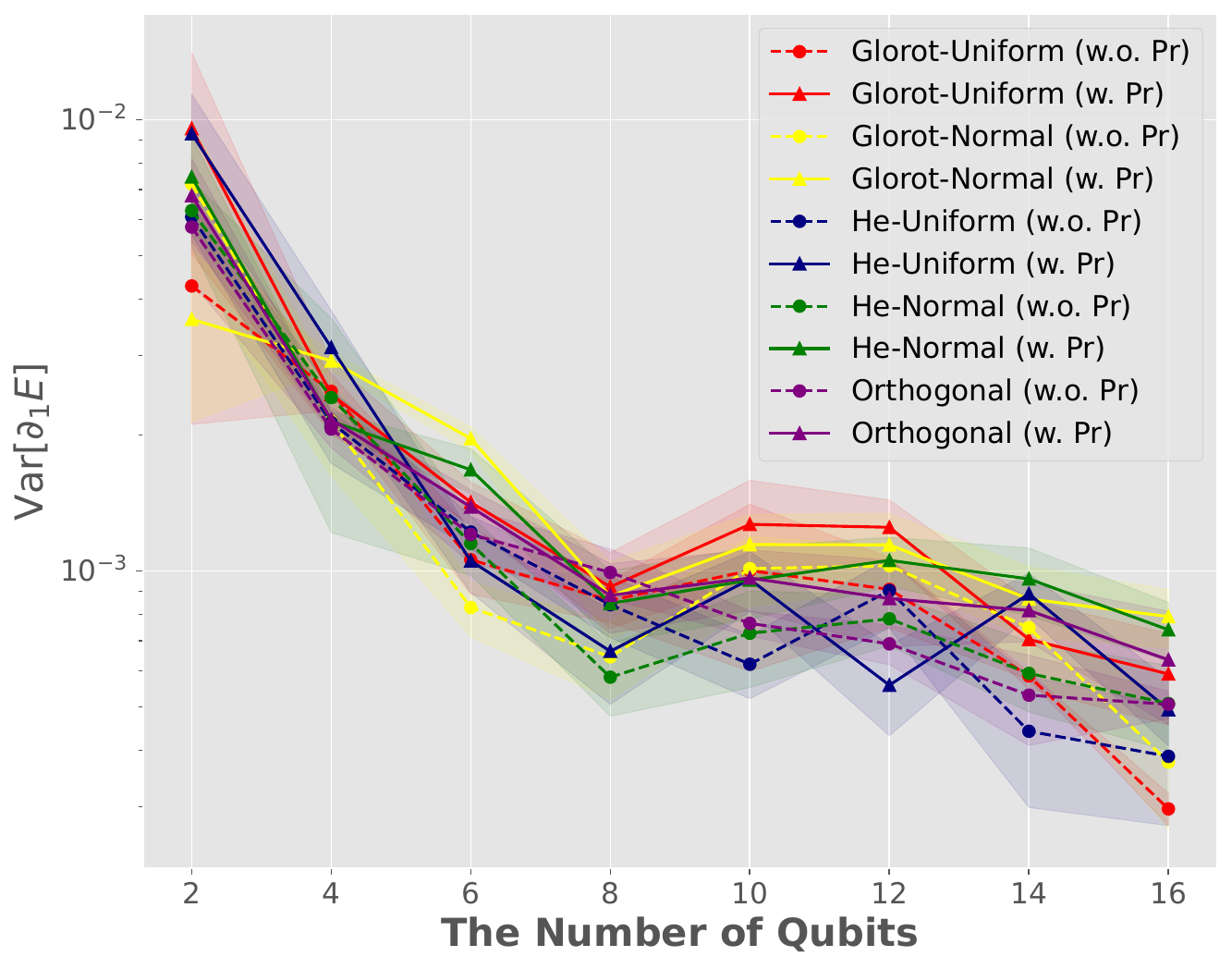}
    \includegraphics[width=0.49\textwidth]{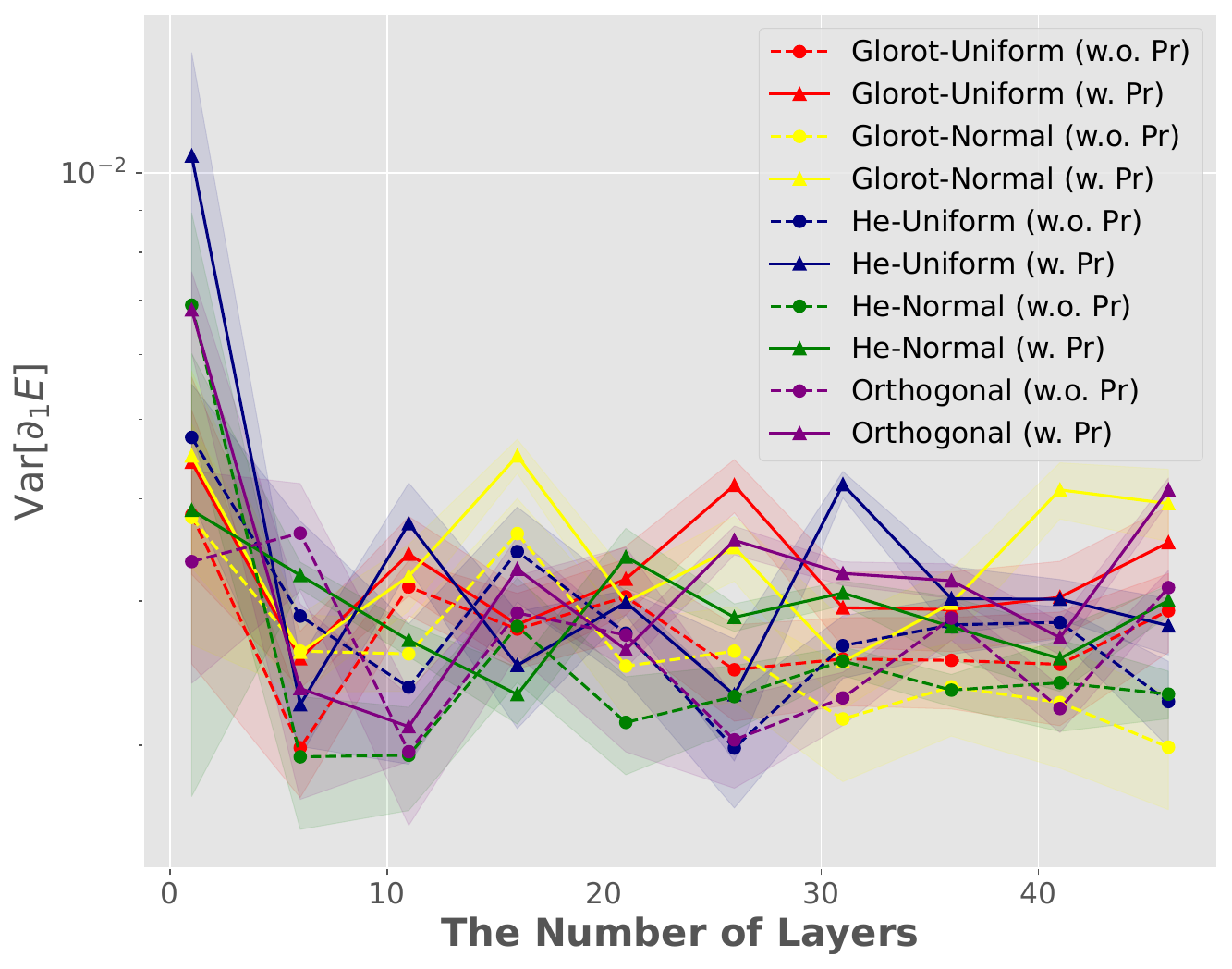}
    \caption{MNIST}
  \end{subfigure}
\caption{Investigation of the mechanism that leverages prior knowledge of the train data on five classic initialization methods. ``w. Pr'' and ``w.o. Pr'' denote whether or not we apply prior knowledge.}
\label{fig:pr_exp2}
\end{figure*}
\paragraph{Validation of the Generalization of Our Proposed Mechanism}
We extend the experiments to validate the generalization of our proposed mechanism on five additional initialization strategies across four public datasets. Before presenting the results, we first introduce these initialization strategies as follows.
\begin{itemize}
 \item {\bf Glorot initialization}~\cite{glorot2010understanding} (a.k.a. Xavier initialization) is a weight initialization method designed to maintain stable signal magnitudes during forward and backward propagation. This method contains two variants. Glorot Uniform method samples the model weights from a uniform distribution bounded by $\sqrt{\frac{6}{{fan_{in} + fan_{out}}}}$, whereas Glorot Normal method samples the model weights from a normal distribution $\mathcal{N} \left(0, \sqrt{\frac{2}{{fan_{in} + fan_{out}}}}\right)$, where $fan_{in}$ ($fan_{out}$) denote the number of input (output) neurons to the layer. The goal of the Glorot initialization method is to balance forward and backward propagation, preventing gradient vanishing or explosion issues.
 \item {\bf He initialization}~\cite{he2015delving} is further improved for mitigating gradient vanishing issues, particularly in deep neural networks using ReLU activation functions. This method has two variants. He Uniform method draws the weights from a uniform distribution bounded by $\sqrt{\frac{6}{fan_{in}}}$, while the He Normal method extracts the weights from a normal distribution $\mathcal{N} \left(0, \sqrt{\frac{2}{fan_{in}}} \right)$.
 \item {\bf Orthogonal initialization}~\cite{hu2020provable} aims to ensure the weight maintains its norm during forward and backward propagation. This method involves initializing weights as an orthogonal matrix, typically obtained via singular value decomposition (SVD) or QR decomposition of a randomly generated matrix. In brief, this method helps preserve gradient magnitudes, making it particularly effective in deep neural networks.
\end{itemize}
We follow the same settings as the experiments presented in Figure~\ref{fig:pr_exp} and observe similar experimental results in Figure~\ref{fig:pr_exp2}. First, it is expected that the gradient variance decreases as the number of qubits or layers increases. Second, applying prior knowledge during initialization can maintain higher gradient variance compared to the cases without it, i.e., the solid lines are higher than the dashed lines. These observations demonstrate the generalization of our approach across five initialization strategies.

\begin{figure}[h]
\centering
  \includegraphics[width=0.9\linewidth]{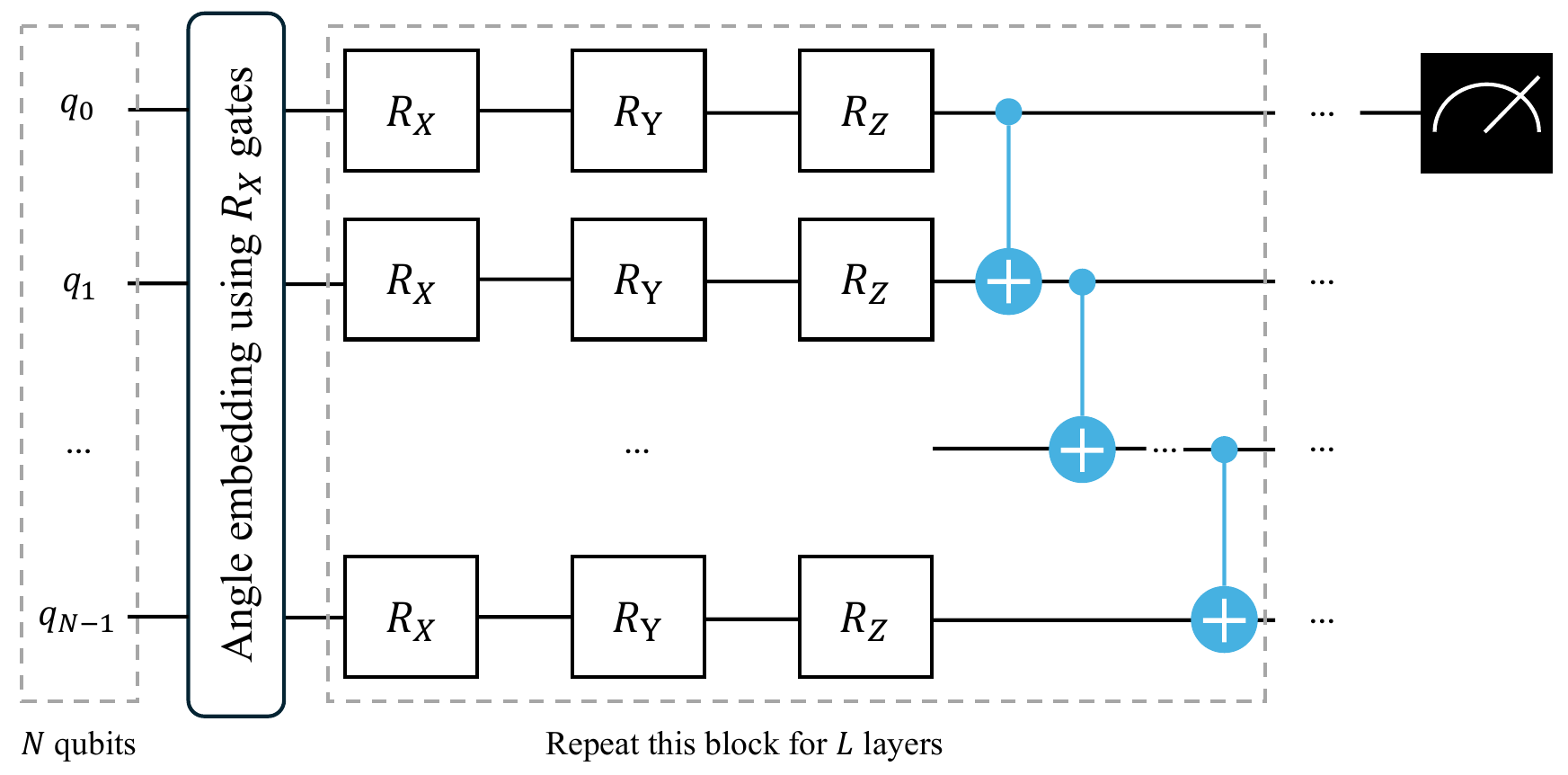}
  \caption{Model architecture of our baseline VQCs.}
\label{fig:vqc}
\end{figure}

\paragraph{Hyper-parameters and Settings}
In this study, we examine our proposed strategy on a baseline VQC, whose model architecture is described in Figure~\ref{fig:vqc}. In particular, we first employ an angle embedding approach to encode classical data into the quantum-state data and further manipulate qubits using a group of parameterized rotation gates. In our baseline VQC, we fix the number of rotation gates to three. In the end, we measure the expectation value on the first qubit of the Pauli-Z gate for binary classification.

\paragraph{Hardware and Software}
The experiment is conducted on a server with the following settings:
\begin{itemize}[itemsep=-1mm]
  \item Operating System: Ubuntu 22.04.3 LTS
  \item CPU: Intel Xeon w5-3433 @ 4.20 GHz
  \item GPU: NVIDIA RTX A6000 48GB
  \item Software: Python 3.11, PyTorch 2.1, Pennylane 0.31.1.
\end{itemize}